\documentclass[10pt,journal,compsoc]{IEEEtran}

\pdfoutput=1
\usepackage[nocompress]{cite}
\usepackage[caption=false,font=footnotesize,labelfont=sf,textfont=sf]{subfig}
\usepackage{url}
\usepackage{pifont}
\usepackage[utf8]{inputenc}
\usepackage{booktabs}
\usepackage{array}
\usepackage{graphicx}
\usepackage{multirow}
\usepackage[table]{xcolor}
\usepackage[flushleft]{threeparttable}
\usepackage{tikz}
\usetikzlibrary{arrows,shapes,positioning,shadows,trees}
\usepackage{enumitem}

\newcolumntype{L}[1]{>{\raggedright\let\newline\\\arraybackslash\hspace{0pt}}m{#1}}
\newcolumntype{C}[1]{>{\centering\let\newline\\\arraybackslash\hspace{0pt}}m{#1}}
\newcolumntype{R}[1]{>{\raggedleft\let\newline\\\arraybackslash\hspace{0pt}}m{#1}}

\newcommand*\rot{\rotatebox{90}}
\newcommand*\OK{\ding{51}}

\begin{document}

\title{A Taxonomy for Management and Optimization \\of Multiple Resources in Edge Computing}

\author{Klervie~Toczé,
        Simin~Nadjm-Tehrani
\IEEEcompsocitemizethanks{\IEEEcompsocthanksitem
 K. Toczé and S. Nadjm-Tehrani are with the Department of Computer and Information Science at Linköping University, Linköping, Sweden.\protect\\

E-mail: [klervie.tocze, simin.nadjm-tehrani]@liu.se
}%
\thanks{}}

\IEEEtitleabstractindextext{
\begin{abstract}
Edge computing is promoted to meet increasing performance needs of data-driven services using computational and storage resources close to the end devices, at the edge of the current network. 
To achieve higher performance in this new paradigm one has to consider how to combine the efficiency of resource usage at all three layers of architecture: end devices, edge devices, and the cloud. While cloud capacity is elastically extendable, end devices and edge devices are to various degrees resource-constrained. Hence, an efficient resource management is essential to make edge computing a reality. 
In this work, we first present terminology and architectures to characterize current works within the field of edge computing. Then, we review a wide range of recent articles and categorize relevant aspects in terms of 4 perspectives: resource type, resource management objective, resource location, and resource use. 
This taxonomy and the ensuing analysis is used to identify some gaps in the existing research. Among several research gaps, we found that research is less prevalent on
data, storage, and energy as a resource, and less extensive
towards the estimation, discovery and sharing objectives. As for resource types, the most well-studied resources are computation and communication resources. Our analysis shows that resource management at the edge requires a deeper understanding of how methods applied at different levels and geared towards different resource types interact. Specifically, the impact of mobility and collaboration schemes requiring incentives are expected to be different in edge architectures compared to the classic cloud solutions. Finally, we find that fewer works are dedicated to the study of non-functional properties or to quantifying the footprint of resource management techniques, including edge-specific means of migrating data and services. 

\end{abstract}

}

\maketitle

\IEEEdisplaynontitleabstractindextext
\IEEEpeerreviewmaketitle

\IEEEraisesectionheading{\section{Introduction}\label{sec:introduction}}

\IEEEPARstart{R}{ecently}, the edge computing paradigm, which consists in having network nodes with computational and storage resources close to the devices (mobile phones, sensors), at the edge of the current network, has attracted interest from both industry and researchers, carrying the promise of a new communication era in which industry can meet the rising performance needs of future applications. 

Indeed, with a forecast of 9 billion mobile subscriptions in the world by 2022, of which 90\% will include mobile broadband, coupled to an eightfold increase in mobile traffic and 17.6 billion of Internet of Things (IoT) devices also sending data \cite{EricssonMobilityReport2017}, there will be a considerable strain put on the network. The current network technologies need to undergo a paradigm shift in order to handle this situation  \cite{EdgeAnalyticsIoT}. Therefore, the aim is to avoid overwhelming the network up to the cloud, and, when possible, move some computing and data analysis closer to the users, to enable better scalability \cite{EmergenceEdge}. Thus, the main idea of edge (or fog) computing is to have intermediate computing facilities between the end devices and the current cloud. As suggested by Amardeep et al. \cite{HowBeneficialIntermediateDC}, this would also enable the current telecom network operators to reduce their operational costs.  

In addition to this, moving computing and storage to the edge of the network has other benefits \cite{EmergenceEdge} such as reducing the latency and jitter \cite{DEVS}, which is especially important for real-time applications such as self-driving cars. Moreover, it enables more privacy for the users by making it possible to keep private data at the edge and enforce privacy policies for the data sent to the cloud (such as blurring sensitive info on a video \cite{EdgeAnalyticsIoT}). Finally, edge networking makes the applications more resilient by being able to process requests at the edge even if the central cloud is down. 

\begin{figure*}[!t]
\centering
\subfloat[Edge server]{\includegraphics[width=1.5in]{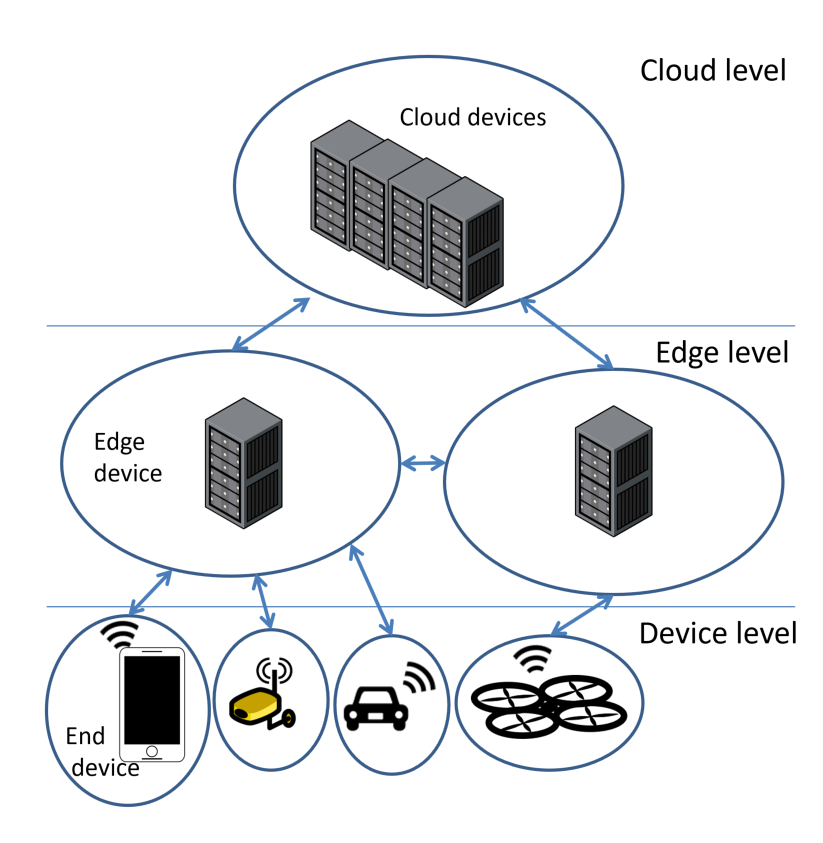}
\label{fig_catA}}
\hfil
\subfloat[Coordinator device]{\includegraphics[width=1.5in]{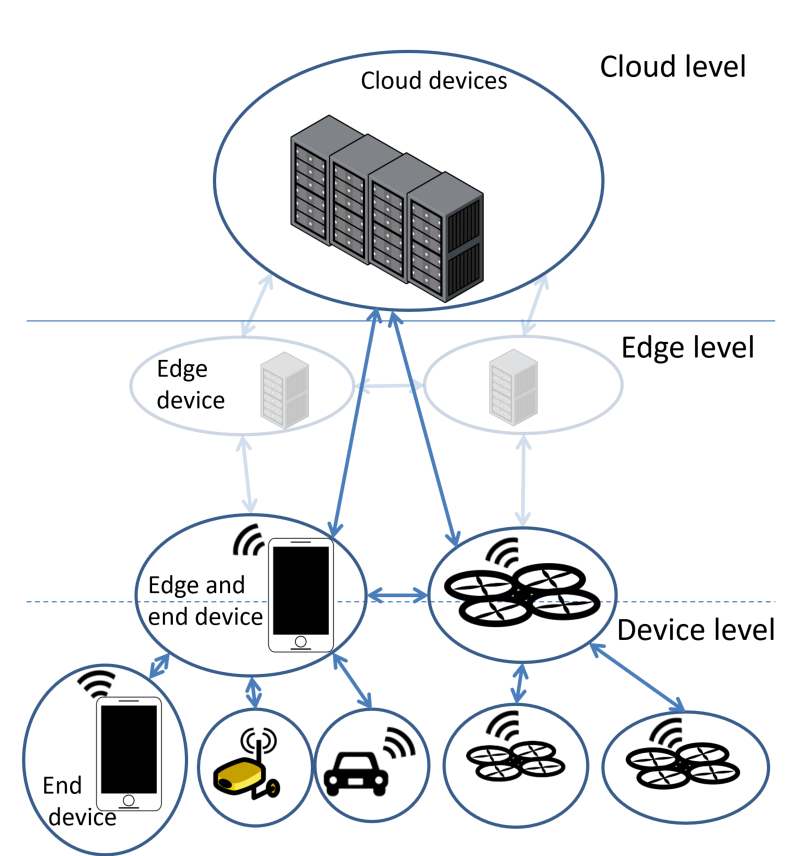}
\label{fig_catB}}
\hfil
\subfloat[Device cloud]{\includegraphics[width=1.5in]{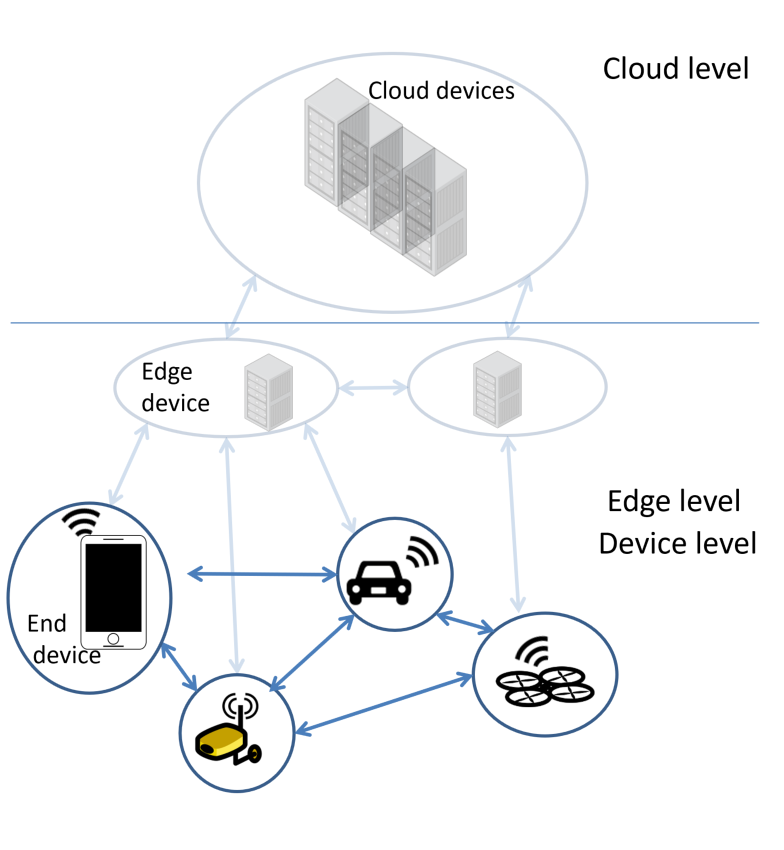}
\label{fig_catC}}
\centering
\caption{Categories of architectures used in edge computing.}
\label{fig_architectures}
\end{figure*} 

In order to achieve this and to make edge computing a reality and a success, there is a need for an efficient resource management at the edge. Indeed, mobile devices or IoT devices are resource-constrained devices, whereas the cloud has almost unlimited but far away resources. Providing and/or managing the resources at the edge will enable the end device to spare resources (e.g. stored energy in batteries), speed up computation, and allows using resources it does not possess. Moreover, keeping data close to where it was generated enables better control, especially for privacy-related issues. Finally, being located close to the user, edge computing makes it possible to increase the quality of provided services through the use of profiling within a local context, without compromising the privacy or having to handle a large number of users. This is known as \textit{context adaptation}.
 
Even though this is still an emerging research area, there is a lot of work ongoing under different denominations including mobile cloud computing \cite{FernandoSurvey}, fog computing \cite{FogReferencePaper}, edge computing \cite{EmergenceEdge}, mobile edge computing \cite{RomanSurvey}, path computing \cite{Mortazavi_CloudPath}, mobile edge cloud \cite{LiuSurvey}, mobile edge network \cite{HowBeneficialIntermediateDC}, infinite cloud \cite{TarnebergResourceManagementChallenges}, follow-me cloud \cite{Taleb_FollowMeCloud}, mobile follow-me cloud \cite{EdgeCachingMobilityPrediction}, multi-tier cloud federations \cite{MultiCloudThesisVamisPaperChapter6}, small cell cloud \cite{Lobillo_SmallCellCloud}, fast moving personal cloud \cite{Wang_FastMovingPersonalCloud}, CONCERT \cite{Liu_CONCERT}, distributed clouds \cite{DistributedCloudSurvey}, and femtoclouds \cite{Habak_FemtoCloud,Habak_WorkloadManagement}. 

Independently of the terminology chosen, which might follow the current naming trend, a common concept here is an intermediate level between the device and the traditional cloud. It is possible to find in the literature numerous surveys about those paradigms in general \cite{FernandoSurvey, AhmedSurvey, StojmenovicSurvey, ChiangSurvey, LiuSurvey, WeisongEdgeVisionChallenges, DataAnalyticsNetworkEdge}, specific aspects of them such as security \cite{RomanSurvey, QunSecurityPrivacyFogSurvey} or specific techniques such as Software-Defined Networking (SDN) \cite{Baktir_Survey}. However, those typically do not consider the resource aspect. The existing surveys about resources either consider it at a high level \cite{YiSurvey} or consider only resource/service provisioning metrics \cite{MahmudSurvey}.

One area that is of high importance and present in many use-cases in edge computing is \textit{offloading}. 
This is the idea of executing a task on another device than the current execution target, typically one having more powerful computational capacities or being less energy-constrained.
Resource management is tightly connected to offloading since in order to take a decision to offload one needs to have knowledge about system resources. 
This knowledge is provided by resource management techniques. For example, resource discovery can be used as an input for taking an offloading decision while resource allocation techniques can be used to perform the offloading decision.
To the best of our knowledge, existing surveys about resource management for offloading at the edge focus on an end device perspective \cite{OffloadingSurvey, OffloadingSurvey2}, on the resource allocation part of resource management \cite{MaoSurveyUnpublished, MachSurvey} or on a single-user/multi-user perspective \cite{MaoSurveyPublished}.

We aim to complement those surveys by providing a more comprehensive perspective. That is, (a) we consider allocation as one among five resource management objective, (b) we consider edge resources in addition to end device or cloud resources, (c) we address multiple types of resources and interrelations amongst them, (d) we review aspects related to locality and what the resource is intended for.

In selecting the survey papers, work considering direct interactions from a device to a cloud \cite{ChunlinCostEnergyAware}, or focusing on cloud performance by offloading to the edge  
\cite{Indices} are not considered. However, offloading between edge devices, or from the edge to the cloud when edge resources are also considered, are included. 
All included papers consider the notion of edge which we attempt to characterize by defining edge-specific architectural instances. This will be done independently of the terminology the authors chose to use.
This paper is a substantial extension of our previous much shorter review \cite{SecureEdgePaper}.

In the remaining parts of this paper, we will first present the terminology used, define edge-specific architectures, and present the proposed taxonomy in Section \ref{sec_architectureTaxonomy}. The taxonomy is then exemplified by an extensive review of papers, which are categorized using the taxonomy elements introduced, namely resource type (Section \ref{sec:resourceType}), resource management objective (Section \ref{sec:Objective}), resource location (Section \ref{sec:ResourceLocation}), and resource use (Section \ref{sec:ResourceUse}). We then discuss research challenges in Section \ref{sec:Challenges} and conclude in Section \ref{sec_conclusion}.

\section{Architectures and research taxonomy}
\label{sec_architectureTaxonomy}

Edge computing is an innovative area bringing together diverse business sectors such as telecommunication actors, vehicle vendors, cloud providers, and emerging application or device providers e.g. for augmented reality. Therefore, the terminology used in research works is diverse and still evolving and multiple architectures are considered. 

In this section, we present first the relevant terminology associated with edge computing that will be used in the rest of the paper. Then, we discuss the current architectures used and present an architectural breakdown that will be the basis for classifying existing research. Finally, we present our proposed research taxonomy and use it to classify the surveyed works. 

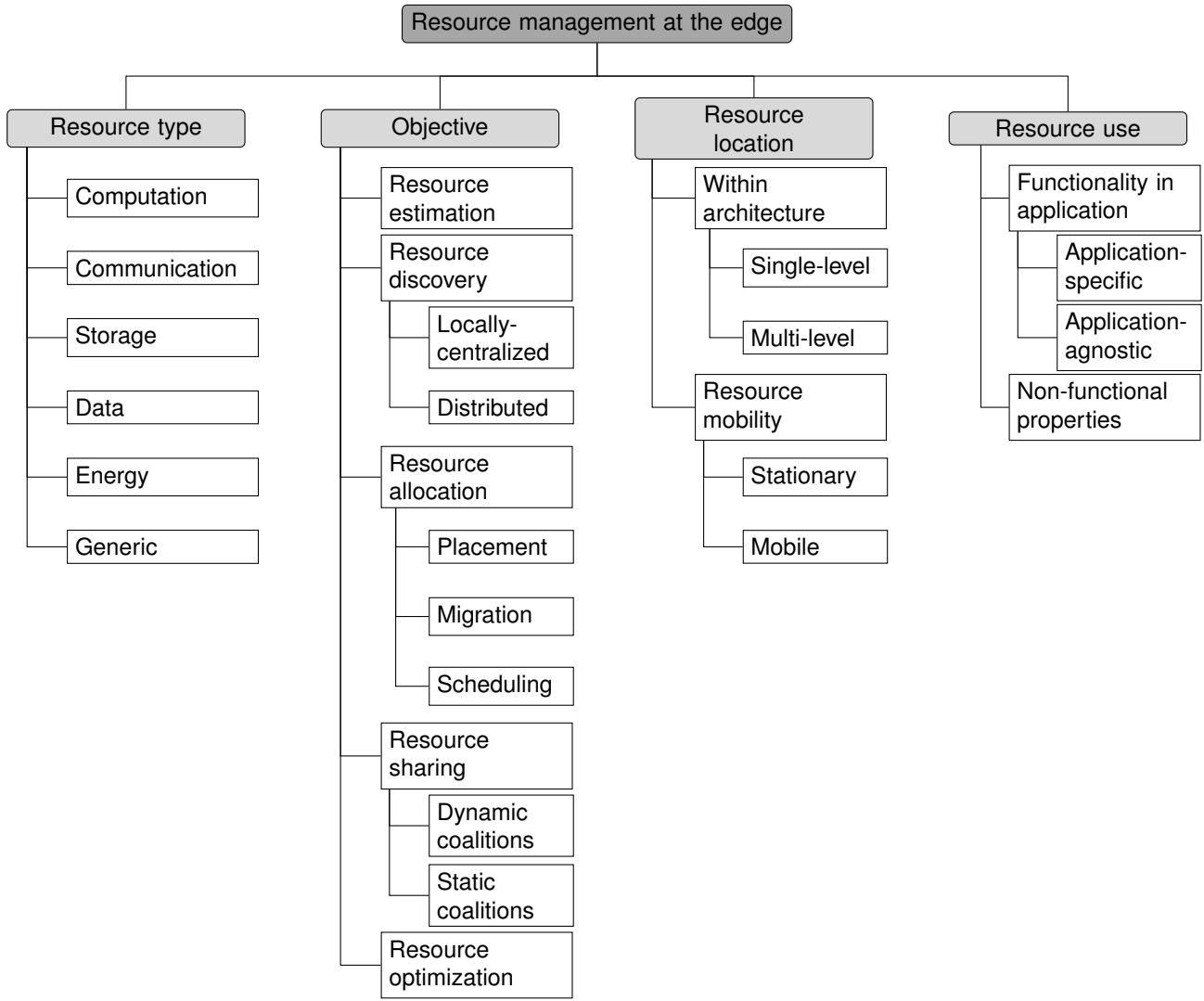
\begin{figure*}[!t]
\centering
\tikzset{
  basic/.style  = {draw, text width=2cm, font=\sffamily, rectangle},
  root/.style   = {basic, rounded corners=2pt, thin, align=center,
                   fill=gray!70, text width=16em},
  level 2/.style = {basic, rounded corners=2pt, thin,align=center, fill=gray!30, text width=9.5em},
  level 3/.style = {basic, thin, align=left, fill=white!60, text width=7.5em, level distance=1.75cm},
  level 4/.style = {basic, thin, align=left, fill=white!60, text width=5.5em}
}

\begin{tikzpicture}[
  level 1/.style={sibling distance=45mm},
  edge from parent/.style={ -,draw},
  edge from parent path={(\tikzparentnode\tikzparentanchor) -- +(0pt,-.5\tikzleveldistance)                                       -| (\tikzchildnode\tikzchildanchor)}]
  >=latex]

\node[root] (c0) {Resource management at the edge}
  child {node[level 2] (c1) {Resource type}}
  child {node[level 2] (c2) {Objective}}
  child {node[level 2] (c3) {Resource \\ location}}
  child {node[level 2] (c4) {Resource use}}
  ;
  
\begin{scope}[every node/.style={level 3}]
\node [below of = c1, xshift=15pt] (c11) {Computation};
\node [below of = c11] (c12) {Communication};
\node [below of = c12] (c13) {Storage};
\node [below of = c13] (c14) {Data};
\node [below of = c14] (c15) {Energy};
\node [below of = c15] (c16) {Generic};

\node [below of = c2, xshift=15pt] (c21) {Resource \\ estimation};
\node [below of = c21] (c22) {Resource \\ discovery};
\end{scope}

\begin{scope}[every node/.style={level 4}]

\node [below of = c22, xshift=10pt] (c221) {Locally-\\ centralized};
\node [below of = c221] (c222) {Distributed};

\end{scope}

\begin{scope}[every node/.style={level 3}]
\node [below of = c222, xshift=-10pt] (c23) {Resource \\ allocation};

\node [below of = c3, xshift=15pt] (c31) {Within \\architecture};

\node [below of = c4, xshift=15pt] (c41) {Functionality in application};

\end{scope}

\begin{scope}[every node/.style={level 4}]

\node [below of = c23, xshift=10pt] (c231) {Placement};
\node [below of = c231] (c232) {Migration};
\node [below of = c232] (c233) {Scheduling};

\node [below of = c31, xshift=10pt] (c311) {Single-level};
\node [below of = c311] (c312) {Multi-level};

\node [below of = c41, xshift=10pt] (c411) {Application-specific};
\node [below of = c411] (c412) {Application-agnostic};

\end{scope}

\begin{scope}[every node/.style={level 3}]
\node [below of = c312, xshift=-10pt] (c32) {Resource \\mobility};
\node [below of = c412, xshift=-10pt] (c42) {Non-functional properties};
\node [below of = c233, xshift=-10pt] (c24) {Resource \\ sharing};
\end{scope}

\begin{scope}[every node/.style={level 4}]

\node [below of = c32, xshift=10pt] (c321) {Stationary};
\node [below of = c321] (c322) {Mobile};

\node [below of = c24, xshift=10pt] (c241) {Dynamic coalitions};
\node [below of = c241] (c242) {Static \\coalitions};

\end{scope}

\begin{scope}[every node/.style={level 3}]
\node [below of = c242, xshift=-10pt] (c25) {Resource \\ optimization};
\end{scope}

\foreach \value in {1,...,6}
  \draw[-] (c1.191) |- (c1\value.west);

\foreach \value in {1,...,5}
  \draw[-] (c2.191) |- (c2\value.west);
  \foreach \value in {1,...,2}
  \draw[-] (c22.201) |- (c22\value.west);
    \foreach \value in {1,...,3}
  \draw[-] (c23.201) |- (c23\value.west);
  \foreach \value in {1,...,2}
  \draw[-] (c24.201) |- (c24\value.west);
  
\foreach \value in {1,2}
  \draw[-] (c3.197) |- (c3\value.west);
  \foreach \value in {1,2}
  \draw[-] (c31.201) |- (c31\value.west);
  \foreach \value in {1,2}
  \draw[-] (c32.201) |- (c32\value.west);

\foreach \value in {1,2}
  \draw[-] (c4.191) |- (c4\value.west); 
  \foreach \value in {1,2}
  \draw[-] (c41.201) |- (c41\value.west);
\end{tikzpicture}
\caption{A taxonomy of resource management at the edge.}
\label{fig_taxonomy}
\end{figure*}

\subsection{Terminology}
\label{sec:sub_terminology}
Following the development of the IoT, it is nowadays not only computers or smartphones which can be connected to the network, but a large variety of things such as cars, sensors, drones, robots, or home appliances. In this survey, all those objects located at the user end of the network, which produce data or need cloud/edge resources will be called \textit{end devices}. 

Devices installed at the edge specifically for edge computing purposes are called \textit{edge devices}. We also include under this term the devices that are already now connecting the end devices to the rest of the network, for example, home routers, gateways, access points, or base stations which are becoming increasingly powerful \cite{RichardContainer}. 

Finally, physical components of the cloud are referred to by the term \textit{cloud devices}. 

We use those network device classifications to create different \textit{levels} in the network: the device level, the edge level, and the cloud level.
Resources which are managed are used to perform \textit{tasks} at some level of the architecture. These can be composed to provide a \textit{service} to the user.

\subsection{Current status of edge architectures}

There is currently no standard architecture for edge computing, although industry and research initiatives exist such as the Open Edge Computing\footnote{\url{http://openedgecomputing.org/}} community, the Open Fog Consortium \footnote{\url{https://www.openfogconsortium.org/}}, and a European Telecommunications Standards Institute (ETSI) standardization group working on Multi-access Edge Computing\footnote{\url{http://www.etsi.org/technologies-clusters/technologies/multi-access-edge-computing}}. Current standardization efforts coming from the ETSI group have been reviewed in detail by Mao et al.\cite{MaoSurveyPublished} and Mach et al. \cite{MachSurvey}. Mao et al.\cite{MaoSurveyPublished} also present edge standardization efforts within the 5G standard.

Therefore, current research on edge computing is using several different architectures and there is ongoing work for defining edge computing architectures. Recent surveys focus on presenting these architectures. For example,  Liu et al. \cite{LiuSurvey} review different architectures for Mobile Edge Cloud servers and networks, and Mach et al. \cite{MachSurvey} present an overview of proposed solutions enabling computation to be brought close to the end device within the field of mobile edge computing. The approach chose by Mouradian et al. \cite{MouradianSurvey} is to classify the architectures depending on whether they are application-specific or not. They also elaborate on architectural challenges according to 6 criteria including scalability and heterogeneity. Our classification of the device types above is consistent with all the surveys on architecture so far. 

\subsection{Used breakdown of architectures} 
\label{sec_archBreak}

In this survey, we choose to classify the different architectures into three main categories inspired by the work of Mtibaa et al. \cite{TowardsMobileOpportunisticComputing} and presented in Figure \ref{fig_architectures}. Those categories are technology-independent and aim at visualizing three high-level variants of the edge computing concept that the current works are using.

The first category, named \textit{Edge server} and depicted in Figure \ref{fig_catA}, is a generic architecture where devices are connected to an edge server, which itself is connected to the rest of the network, including the cloud. In this type of architecture, the edge server is at a fixed physical location and has relatively high computational power, though it remains less powerful than a conventional data center used in the cloud computing paradigm. Moreover, there is a clear separation between the device level and the edge level. 
In the literature, such edge servers are named for example cloudlets\cite{CloudletReferencePaper, CSCAN}, micro data centers \cite{PREFog, Singh_RTSANE}, nano data centers \cite{GreeningNanoDataCenters} or local cloud \cite{ResourecAllocSchemeVehicular}. 
They can be located for example in shops, enterprises, or co-located with the base stations of the telecom access network. Indeed, in the ongoing work on what the fifth generation (5G) of telecommunication networks will look like, a cloud radio access network (C-RAN) is envisaged \cite{5GSurvey, FinedGrainedResourceAware}, with connections to other edge computing areas such as mobile cloud computing \cite{CostEffectiveResourceAllocaion}.

The second category, named \textit{Coordinator device} and depicted in Figure \ref{fig_catB}, is an architecture where one end device acts as a coordinator between the other end devices. It also acts as a proxy towards an edge device and/or the cloud if such connectivity is needed. The difference between a coordinator device and an edge server is that the coordinator device can be mobile and has less computational power and bandwidth than an edge server. In this architecture category, the border between the device level and the edge level is not a sharp one, as the coordinator level providing edge functionality is actually an end device. 
Solutions using this category of architecture are named for example  fog colonies with a control node  \cite{TowardsQoSAwareFogServicePlacement}, vehicular clouds with a cluster head \cite{ClusterBasedVehicularClouds} and local clouds with a local resource coordinator \cite{ServiceOrientedHeterogeneousResourceSharing}. It is interesting to note here that the term \textit{local cloud}, which was already used for describing a part of the edge server architecture category described in the previous paragraph, is used to describe various architectural solutions, illustrating well the fact that the terminology used in edge computing is not yet set.

The last category, named \textit{Device cloud} and depicted in Figure \ref{fig_catC}, is an architecture where the end devices communicate with each other to find needed resources and deliver the wanted services. The devices might communicate with an edge device connected to the cloud if needed but this is not necessary. In this architecture category, the device level and the edge level are thus merged.
Research work considering this category of architecture call it opportunistic computing \cite{ServiceProvisioningOpportunistic},  cooperation-based mobile cloud computing \cite{AdaptiveResourceDiscovery, OpportunisticResourceSharingMCC}, or transient clouds \cite{TransientCloud}. 

While all these architectures need to be populated with dedicated resource management elements there is no general agreement about where to place the needed policies. A recent proposal for a generic software architecture that encompasses the edge server version in figure~\ref{fig_catA} is an enabler for evaluation of multiple resource management policies within common testbeds~\cite{WMFOG}. 

\subsection{Taxonomy of edge resource management}

In addition to classifying the reviewed papers according to the architecture category they consider, we also present a taxonomy of resource management at the edge. This taxonomy, illustrated in Figure \ref{fig_taxonomy}, aims at getting an overview of state-of-the-art research in this area and presents four main aspects: resource type, objective of resource management, resource location, and resource use. 

The two first aspects were constructed by reviewing the current \textit{type} of resources used and the \textit{objective} for which they are used in the literature. The two last aspects are based on mutually-exclusive pairs for describing the resource \textit{location} and the \textit{use} of the resource.  

In the coming sections, we will describe the different parts of the taxonomy, and how the surveyed works can be placed in the four above contexts, as well as the architectural models described.  

\section{Resource type}
\label{sec:resourceType}
The first step in evaluating the benefit of an edge solution is to decide what are the resource types that can be managed in a better way compared to a centralized system. 

An obvious justification for using edge architectures is reducing the response time, which can be done if computation and communication resources are provided and used adequately. Storage as a resource is also a concern since local storage may benefit security or timeliness due to customized fetching and secure storing mechanisms. A less obvious type of resource is having access to a special type of data (e.g. availability of sensors) that provides local benefits in an application. Examples are the use of cameras or location sensors. The amount and type of data captured in turn affects computation and communication resources (how often to shuffle data, how much to process or filter before shuffling), and implicitly the choice of where and how much of other resources to deploy. The fifth category we consider is energy as a resource, which is clearly influenced by the amount of computation, communication, storage, and data capturing that goes on. Finally, some works consider resources in a generic way using abstract terms such as "Virtual Resource Value" or just as unit-less elements in a model.

Table \ref{table_articles_types} summarizes the surveyed papers in terms of their mapping to the architectural choices in Figure \ref{fig_architectures}. It also shows which resource is focused on within each work, either specific or generic. As it can be seen, the vast majority of the surveyed articles focus on several resources. Therefore, this section will present the common combinations of resources described above and presented in Figure \ref{fig_taxonomy}.  

\begin{table*}
\centering
\caption{Surveyed articles according to architecture category from Figure \ref{fig_architectures} and resource type.} 
\label{table_articles_types}
\begin{tabular}{@{} clcccccc @{}} 
 & Article & Computation & Communication & Storage & Data & Energy & Generic \\ 
\midrule
\multirow {23} {*} {\rot{Edge server}}  & Liu \cite{AdaptiveMultiResourceAllocationCloudlet} & \OK & \OK & & & &\\
& Confais \cite{ObjectStore} &  & \OK& \OK &  & &\\
& Aazam \cite{PREFog} & & & & & & \OK\\ 
& Arkian \cite{MIST}  &\OK & \OK& \OK& & &\\
& Aazam \cite{FogDynamicResourceEstimation}   & & & & & & \OK\\ 
& Fan \cite{EnergyDrivenAvatar} & \OK& & & &\OK &\\
& Oueis \cite{OueisThesis} & \OK& \OK& & & \OK&\\
& Tang \cite{DoubleSidedBidding} & \OK& \OK& & & &\\
& Borylo \cite{EnergyAwareFogAndCloud} & & & & &\OK &\\
& Yousaf \cite{FinedGrainedResourceAware} & \OK&\OK &\OK & & &\\
& Wang \cite{CostEffectiveResourceAllocaion} & \OK & \OK& & & &\\
& Gu \cite{CostEfficient}      & \OK& \OK&\OK & & &\\
& Tärneberg \cite{DynamicApplicationPlacement}  & \OK & \OK& & & &\\
& Plachy \cite{DynamicResourceAlllocation} &\OK & \OK& & & &\\
& Gomes \cite{EdgeCachingMobilityPrediction} & & & & \OK& &\\
& Fricker \cite{OffloadingSchemeDCFog} &\OK & & & & &\\
& Rodrigues \cite{HybridMethodServiceDelay} & \OK& \OK& & & &\\
& Zhang \cite{WeisongFirework} &\OK & & &\OK & &\\
& Bittencourt \cite{MobilityAwareApplicationScheduling}  & \OK& \OK& & & &\\
& Zamani \cite{DeadlineConstrainedVideoAnalysis} &\OK & \OK & & & &\\
& Valancius \cite{GreeningNanoDataCenters} & & \OK&\OK & &\OK &\\
& Chen \cite{Chen_sociallyTrusted} &\OK &\OK & & &\OK &\\
& Wang \cite{Wang_ElasticUrbanVideosur} & \OK& \OK & \OK&& &\\ 
& Yi \cite{Yi_LAVEA} & \OK& \OK& & & &\\
& Wang \cite{Wang_DynamicServicePlacementJournal} & & & & & &\OK\\
& Sardellitti \cite{Sardellitti_JointOptimization} &\OK &\OK & & &\OK &\\
& Singh \cite{Singh_RTSANE} &\OK &\OK & & & &\\
\midrule
\multirow {7} {*} {\rot{\parbox{1.5cm}{\centering Coordinator \\ device}}}   & Nishio \cite{ServiceOrientedHeterogeneousResourceSharing}   &\OK & \OK& & \OK & \OK &\\
& Skarlat \cite{TowardsQoSAwareFogServicePlacement} &\OK &\OK & \OK& \OK & &\\
&Borgia \cite{MobileEdgeCloudsForInformationCentric}  & & \OK& &\OK & &\OK\\ 
& Athwani \cite{ResourceDiscoveryInMCC} &\OK &\OK & & & \OK&\\
& Arkian \cite{ClusterBasedVehicularClouds} &\OK &\OK &\OK & & &\\
& Penner \cite{TransientCloud} & & & & & & \OK\\
& Bianzino \cite{Bianzino2014} & & \OK & & & \OK &\\
& Habak \cite{Habak_WorkloadManagement} & \OK &\OK & & \OK& &\\
\midrule
\multirow {7} {*} {\rot{Device cloud}} & Liu \cite{AdaptiveResourceDiscovery} & & & & & \OK& \OK\\
& Mascitti \cite{ServiceProvisioningOpportunistic}&\OK &\OK & & & &\\
& Liu \cite{OpportunisticResourceSharingMCC} & & \OK& & & &\OK\\
& Meng \cite{ResourecAllocSchemeVehicular}    &\OK & \OK& & & &\\
& Qi \cite{DynamicResourceOrchestration}  & & & & &\OK &\OK\\
& Mtibaa \cite{TowardsResourceSharing}   & & & & &\OK&\\
\bottomrule
\end{tabular}
\end{table*}

\subsection{Single resource focus}
Even though the majority of the surveyed papers choose to focus on several resources, some papers focus on only one resource type. We present those papers in this subsection and then move on to multi-resource cases.
\subsubsection{Generic}

When focusing on a single resource type, most of the works use a generic one, that is used as an abstraction for actual resources. 

The abstraction used varies in various articles. For example,  
Penner et al. \cite{TransientCloud} work with device capabilities as an abstraction when proposing resource assignment algorithms. Other works, such as Aazam et al. \cite{PREFog,FogDynamicResourceEstimation}, define a new conceptual unit. "Virtual Resource Value" is the unit for any resource, which is then mapped to physical resources according to the type of service and current policies of the cloud service provider. 

Sometimes the abstraction is at an even higher level: Wang et al. \cite{Wang_DynamicServicePlacementJournal} use generic cost functions that can be used to model many aspects of performance such as monetary cost, service access latency, amount of processing resource consumption or a combination of these.  When proposing a method for online service placement, they however analyze its performance for a subset of cost functions related to resource consumption 
with the claim that this subset is still general. 
 
\subsubsection{Energy}

Some works focus solely on energy, which is especially important at the edge since devices, in particular end devices, are often resource-constrained. 
For example, Mtibaa et al. \cite{TowardsResourceSharing} perform offloading between end devices in order to maximize the group lifetime. 

Still considering only energy but with another perspective, 
Borylo et al. \cite{EnergyAwareFogAndCloud} classify datacenters in two categories (green and brown depending on which source of energy they use) and then use a latency-aware policy to choose a data center for serving a request.

\subsubsection{Other}
\label{sec:singleF_computation}

There are  works that consider a minimum computational resource unit per device. For example, Fricker et al. \cite{OffloadingSchemeDCFog} use servers as an abstraction (one request occupies one server).

Data as a resource, in addition to sensor data mentioned earlier, can also be seen as content. Gomes et al. \cite{EdgeCachingMobilityPrediction}  propose an algorithm for content migration at the edge, together with mobility prediction as an enabler within their new Mobile Follow-Me Cloud architecture. This work builds upon the initial Follow-Me Cloud proposal by Taleb et al. \cite{Taleb_FollowMeCloud}.

\subsection{Multiple resource focus}

All other surveyed articles are focusing on multiple resource types. In this section, we group the papers according to the different combinations of resources they consider.

\subsubsection{Computation and communication}

The most common combination of resource types studied is computational and communicational resources together. Thus, we begin by considering works that study this combination, and in one case together with data.

Liu et al. \cite{AdaptiveMultiResourceAllocationCloudlet} consider wireless bandwidth and computing resource when deciding whether to handle a request in a cloudlet or in the cloud. Another example is the work by Bittencourt et al. \cite{MobilityAwareApplicationScheduling}, who consider  bandwidth between the cloud and cloudlet, as well as cloudlet processing capabilities when evaluating different scheduling strategies.

Computational resources can be addressed at a physical level, e.g. discussing CPU cycles, or at a conceptual level, e.g. use of virtual machines (VMs) as resource elements. In the surveyed articles, Wang et al. \cite{CostEffectiveResourceAllocaion} consider CPU cycles, Singh et al. \cite{Singh_RTSANE} consider Millions of Instructions per Second (MIPS), and Rodrigues et al. \cite{HybridMethodServiceDelay}  consider the number of processors per cloudlet. At a conceptual level, Zamani et al. \cite{DeadlineConstrainedVideoAnalysis} consider different computing resources based on the average number of tasks completed per unit of time,  and Plachy et al. \cite{DynamicResourceAlllocation} allocate computational resources in the form of VMs. 

Sometimes the VMs are used as a means to ensure that a task can run given enough underlying resources in the device hosting the VM, for example in the work by Tärneberg et al. \cite{DynamicApplicationPlacement}. 

Instead of using VMs, Yi et al. \cite{Yi_LAVEA} adopt lightweight OS-level virtualization and a container technique,  
 arguing that resource isolation can be provided at a much lower cost using OS-level virtualization. They also pinpoint that the creation and destruction of container instances is much faster and thus enable the deployment of an edge computing platform with minimal efforts.

As in Section \ref{sec:singleF_computation}, some works consider a minimum resource unit that corresponds to a device. For example, Meng et al. \cite{ResourecAllocSchemeVehicular} consider one vehicle as the minimal computing resource unit. Vehicles are aggregated in a resource pool together with communication resources and resource units from the cloud and the edge.

Communication power needed can be considered as a part of the cost when sharing resources \cite{DoubleSidedBidding}. 
In contrast, communication can be characterized by a delay term impacting the task completion time, like \cite{DeadlineConstrainedVideoAnalysis, ServiceProvisioningOpportunistic,Singh_RTSANE}.

Finally, Habak et al \cite{Habak_WorkloadManagement} consider computation, communication, and data in femtoclouds. The data considered gives information about task dependencies in order to determine in which order the tasks need to be executed and which ones can be run in parallel.

\subsubsection{Computation, communication, and storage}

Other works, in addition to the computation and communication resource types, also include storage in their study.

For example, Arkian et al. \cite{ClusterBasedVehicularClouds} tackle resource issues in vehicular clouds by considering all three resource types. Elsewhere,  crowdsensing is tackled with the same resource considerations~\cite{MIST}. 

Another example is the work by Skarlat et al. \cite{TowardsQoSAwareFogServicePlacement}, where they model service demands, and a specific kind of resource (sensor data) as well as the computational and storage resources. In this work, communication is considered as a delay term.

VMs can also be considered as an encapsulation of the above three resources, in methods that ensure the underlying resources in the device hosting the VM are adequate \cite{CostEfficient}. 

Still considering virtualization, Wang et al. \cite{Wang_ElasticUrbanVideosur} propose a system architecture where applications' requests contain computing complexity and storage space requirements. Those requirements are then translated by a SDN controller node into computing power requirements, bandwidth volumes or requirements on  security groups. When trying to allocate more computing and bandwidth resources in an emergency situation, their system will do it by creating new VMs. 

Finally, in addition to considering computation, communication, and storage, Yousaf et al. \cite{FinedGrainedResourceAware} emphasize the fact that different resources should not be considered in isolation as there are interactions between them. Thus, they describe and use the concept of resource affinity in their scheme.
\subsubsection{Computation, communication, and energy}

Another combination studied by several of the surveyed articles is computation, communication, and energy resource types. 

Athwani et al. \cite{ResourceDiscoveryInMCC} aim at making resource discovery energy-efficient in order to save battery.  Nishio et al. \cite{ServiceOrientedHeterogeneousResourceSharing} consider energy efficiency in their algorithms, but at a more general level, without battery life considerations. 

Oueis \cite{OueisThesis} focus on energy-efficient communication with the aim of minimizing the communication power needed. Similarly, when studying edge collaboration in ultra-dense small base stations networks with trust considerations, Chen et al. \cite{Chen_sociallyTrusted} consider computing (CPU cycles per second), communication as radio-access provisioning,  
and energy used both for transmission and computation. 

Sardellitti et al. \cite{Sardellitti_JointOptimization} propose an algorithmic framework to solve the joint optimization problem of radio
and computational 
resources with the aim of minimizing the overall energy consumption of the users while meeting latency constraints. They first present a solution for the single-user case and then consider the case of offloading with multiple cells, in a centralized and a distributed manner. 

 When considering energy as a resource, a comprehensive discussion of interactions between multiple actions is mapped to energy apportionment policies by Vergara et al. ~\cite{TOMPECS16}. However, since this work considers edge-/cloud-specific apportionments as one among many application areas, i.e. addresses energy sharing in a much wider context, we do not further consider it in our classifications.

\subsubsection{Combinations including generic resources}

We now consider generic resources in association with other resource types, such as energy or communication.

First, Liu et al. \cite{AdaptiveResourceDiscovery} consider abstract tasks and resources to address energy efficiency. They switch between a centralized or a flooding mode depending on energy consumption while keeping the expected value of resource information availability, which is their quality metric. 
Qi et al. \cite{DynamicResourceOrchestration} choose to abstract resources as services and consider energy consumption in the end device when taking an offloading decision.

Regarding communication, Liu et al. \cite{OpportunisticResourceSharingMCC} use the notion of generic resource (when referring to a combination of bandwidth and CPU available for sharing), as well as concrete bandwidth when nodes are at contact range. Borgia et al. \cite{MobileEdgeCloudsForInformationCentric} consider data-centric service providers having storage, computing and networking capabilities, but in their evaluation abstract away the storage and computing resources by only considering the extent to which services are waiting for resources on the provider side. 

\subsubsection{Other combinations}

Not all works considering computation also consider communication. Less common combinations including computational resources are those with energy and data. 

With regards to energy, Fan et al. \cite{EnergyDrivenAvatar} present a virtual machine migration scheme which aims at using as much green energy as possible in the context of green cloudlet networks. 

Data and computation is the focus of Zhang et al. \cite{WeisongFirework}, who studied distributed data sharing and processing in order to use data coming from different stakeholders for new IoT applications and propose a new computing paradigm called Firework.

Less common combinations including communication resources include storage and energy resource types. 

Confais et al. \cite{ObjectStore} present how a storage service can be provided for fog/edge infrastructure, based on the InterPlanetary file system, and scale-out network-attached systems. Their aim is to propose a service similar to the Amazon Simple Storage Service solution\footnote{\url{https://aws.amazon.com/fr/s3/}} for the edge.

Adding energy to storage and communications resources, Valancius et al. \cite{GreeningNanoDataCenters} consider energy-efficient algorithms when introducing a new distributed data center infrastructure for delivering Internet content and services.

Finally, Bianzino et al. ~\cite{Bianzino2014} study the trade-off between bandwidth and energy consumption when an end device has access to multiple networking interfaces and can switch between them. They aim for energy efficiency but use an abstract model of power usage based on the amount of data being shuffled.

\subsection{Summary of resource and architecture choices}

In this section, we have presented the surveyed articles depending on their resource focus. Examining the collection of papers above, resource studies so far seem to focus on computation and communication resources to a greater extent. Moreover, data as a resource is a potential not extensively explored. Similarly, energy is underrepresented among resources studied. 

Furthermore, it is noticeable that storage is not the main focus of attention. It could be due to the fact that the cloud is available as a fall-back in many cases.  It could also be the case that persistent data storage is not the main focus of most of the applications considered at the edge. Rather, the service or completed task is the main purpose. Another reason could be that presently there are not many critical use cases with latency-constrained storage, but this may change when more and more IoT devices appear in the field. An alternative explanation could be that the authors choose to focus on a reduced set of resources for ease of presentation thinking that the work can be extended to other resources such as storage. Such claims, however, have to be considered with care as this is ignoring the fact that there could be interactions between resources as studied by Yousaf et al. \cite{FinedGrainedResourceAware}. 

Some resources are dealt with mainly as physical elements whereas others naturally lend themselves to be defined in abstract ways. For example, sensors are present in the end devices, which can produce useful data needed for the completion of the task (as in \cite{TransientCloud, ServiceOrientedHeterogeneousResourceSharing, WeisongFirework,TowardsQoSAwareFogServicePlacement}), whereas bandwidth (throughput) is a natural abstraction for distinguishing between different radio interfaces or different physical environments (abstracting the impacts of reduced signal strength, interference, etc). 

Moreover, when using a generic resource representation, it is easier to combine several resource types or to combine resources with other performance-related considerations one example being the generic cost function in the work by Wang et al. \cite{Wang_DynamicServicePlacementJournal}. In their performance analysis, they define local and migration resource consumption that can be related for example to CPU and bandwidth occupation or the sum of them.  

Another point to note is that the first architectural instance (Edge server) is the most predominant structure used in the surveyed papers.

\begin{table*}
\centering
\caption{Surveyed articles according to architecture category from Figure \ref{fig_architectures} and objective of resource management.} 
\label{table_articles_objective}
\begin{tabular}{@{} clccccc @{}} 
\cmidrule(r){3-7}
 & & \multicolumn{5}{c}{Objective} \\ 
 \cmidrule(r){3-7} 
 &  & \parbox{1cm}{\centering Resource\\estimation} & \parbox{1cm}{\centering Resource\\discovery} & \parbox{1cm}{\centering Resource\\allocation} & \parbox{1cm}{\centering Resource\\sharing} & \parbox{1cm}{\centering Resource\\optimization} \\ 
\midrule
\multirow {23} {*} {\rot{Edge server}}  & Liu \cite{AdaptiveMultiResourceAllocationCloudlet} &&&\OK&&\OK \\
& Confais \cite{ObjectStore}      &&&\OK&& \\
& Aazam \cite{PREFog}      & \OK &&&& \\
& Arkian \cite{MIST}      &&&\OK&&\OK \\
& Aazam \cite{FogDynamicResourceEstimation}    & \OK &&&& \\
& Fan \cite{EnergyDrivenAvatar}      &&&\OK&&\OK \\
& Oueis \cite{OueisThesis}      &&&\OK&&\OK\\
& Tang \cite{DoubleSidedBidding}      & &&&\OK \\
& Borylo \cite{EnergyAwareFogAndCloud}      &&&\OK&& \\
& Yousaf \cite{FinedGrainedResourceAware}      &&&\OK&&\OK \\
& Wang \cite{CostEffectiveResourceAllocaion}      &&&\OK&&\OK \\
& Gu \cite{CostEfficient}      &&&\OK&&\OK \\
& Tärneberg \cite{DynamicApplicationPlacement} &&&\OK&&\OK \\
& Plachy \cite{DynamicResourceAlllocation}      &&&\OK&&\\
& Gomes \cite{EdgeCachingMobilityPrediction}      &&&\OK&& \\
& Fricker \cite{OffloadingSchemeDCFog}      &&&\OK&& \\
& Rodrigues \cite{HybridMethodServiceDelay}      &&&\OK&&\OK \\
& Zhang \cite{WeisongFirework}      &&&&\OK& \\
& Bittencourt \cite{MobilityAwareApplicationScheduling}      &&&\OK&& \\
& Zamani \cite{DeadlineConstrainedVideoAnalysis}      &&\OK&\OK&&\OK \\
& Valancius \cite{GreeningNanoDataCenters}   &&&\OK&&\OK   \\
& Chen \cite{Chen_sociallyTrusted} &&&&\OK&\\
& Wang \cite{Wang_ElasticUrbanVideosur} &&&\OK&&\\ 
& Yi \cite{Yi_LAVEA} &&&\OK&&\OK\\ 
& Wang \cite{Wang_DynamicServicePlacementJournal} &\OK&&\OK&&\OK\\
& Sardellitti \cite{Sardellitti_JointOptimization} &&&\OK&&\OK\\
& Singh \cite{Singh_RTSANE} &&&\OK&&\\
\midrule
\multirow {7} {*} {\rot{\parbox{1.5cm}{\centering Coordinator \\ device}}}   & Nishio \cite{ServiceOrientedHeterogeneousResourceSharing}   & & &&\OK &\OK \\
& Skarlat \cite{TowardsQoSAwareFogServicePlacement} & &&\OK&\OK&\OK\\
&Borgia \cite{MobileEdgeCloudsForInformationCentric}     &&&\OK&& \\ 
& Athwani \cite{ResourceDiscoveryInMCC}      & &\OK&&\OK&\OK \\
& Arkian \cite{ClusterBasedVehicularClouds}      & &\OK&&\OK&\OK \\
& Penner \cite{TransientCloud}      &&&\OK&& \\
& Bianzino \cite{Bianzino2014} &&&&\OK&\OK\\
& Habak \cite{Habak_WorkloadManagement} &\OK&&\OK&\OK&\OK\\
\midrule
\multirow {7} {*} {\rot{Device cloud}} & Liu \cite{AdaptiveResourceDiscovery} & \OK & \OK && & \OK \\
& Mascitti \cite{ServiceProvisioningOpportunistic} &&&\OK&& \\
& Liu \cite{OpportunisticResourceSharingMCC}    & &&&\OK&\OK \\
& Meng \cite{ResourecAllocSchemeVehicular}    &&&\OK&&\OK\\
& Qi \cite{DynamicResourceOrchestration}      &&&\OK&&\OK \\
& Mtibaa \cite{TowardsResourceSharing}      &\OK &&\OK&\OK&\OK \\
\bottomrule
\end{tabular}
\end{table*}

\section{Objective}
\label{sec:Objective}

A major classification represented in this taxonomy is the objective of resource management. Resource management at the edge can be decomposed into several areas addressing different problems, as shown in the branches under objective in Figure \ref{fig_taxonomy}. In Table \ref{table_articles_objective}, we present which surveyed article addresses which problem(s) and we describe those problems in the following subsections. 
As it can be seen in the table, one surveyed work can address several of the areas.

The resource management objective is orthogonal to the resource types presented in Section \ref{sec:resourceType} but a discussion of the relationship between objectives of resource management and resource types is conducted in our summary in  Section \ref{sec:SummaryObjective}. 

\subsection{Resource estimation}

One of the first requirements in resource management is the ability to estimate how many resources will be needed to complete a task or to carry a load. This is important, especially for being able to handle fluctuations in resource demand while maintaining a good quality of service (QoS) for the user. On the supply side, resources at the edge can be mobile, and thus become unreachable, which makes them less reliable than resources in a data center. On the demand side, user mobility implies that there can be sudden user churn, with the corresponding dynamic requests having to be handled by the edge.

In their work, Liu et al. \cite{AdaptiveResourceDiscovery} use the average of historical data
in order to predict the characteristics of resource distribution and usage for the next time slot. 
The term fog is used by Aazam et al. \cite{PREFog}, who propose that it can be used to perform  future resource consumption estimation as a first step for allocating resources in advance.  They formulate an estimation mechanism which takes into account the reliability of the customer, using what they call the relinquish probability. In another article, Aazam et al. \cite{FogDynamicResourceEstimation}, present the same idea but with an emphasis on how different customers can be charged for the service. Another work by Mtibaa et al. \cite{TowardsResourceSharing} estimates power consumption in order to maximize device lifetime.

Wang et al. \cite{Wang_DynamicServicePlacementJournal} use a look-ahead window for prediction into the future in order to minimize cost over time. They study the optimal size for such a window and propose an algorithm using binary search to find this size which they evaluate as accurate as it gives results close to the size giving the lowest cost. However, the actual prediction mechanism is assumed to be available.  

With respect to computational resources, Habak et al. \cite{Habak_WorkloadManagement} are estimating the task requirements within a job analyzer. They evaluate the sensitivity of their mechanisms to estimation errors and find that the pipeline job model is insensitive to such errors whereas the general parallel path model starts exhibiting a significant increase of job completion time if the estimation error variance exceeds 30\%. 

There are of course many earlier works that use sophisticated prediction mechanisms for estimating future loads in cloud environments (e.g. ~\cite{Xiao2013}) but our focus has been on edge-related papers and instances of estimation therein.

\subsection{Resource discovery}

As opposed to the estimation problem which relates to the demand side, resource discovery is about the supply side. A management system needs to know which resources are available for use, where they are located and how long they will be available for use (especially if the resource providing device is moving or it is battery-driven). This area is especially important at the edge where every resource is not under the control of the system at all times, so the supply is volatile.

The collaboration at the edge can take the form of clusters, as advocated by Atwhani et al.  \cite{ResourceDiscoveryInMCC}. They present an algorithm for forming clusters of devices and performing resource discovery within the cluster. Their strategy is that each member of the cluster will inform the cluster head about their available resources and all requests for resources are handled by the cluster head. 
From their evaluation with respect to energy consumption and delay, they conclude that maintaining the cluster consumes extra energy, especially if the devices are very mobile. 
Arkian et al. \cite{ClusterBasedVehicularClouds} also present a solution using clusters and an algorithm for selecting the cluster head, i.e. the vehicle which will be responsible for maintaining the vehicular cloud resources. They use fuzzy logic and a reinforcement learning technique. In order to select the best vehicle, they need to know which vehicle possesses the best communication to the edge node located on the road-side, hence performing resource discovery. This is done in a similar way to earlier work \cite{ResourceDiscoveryInMCC}, i.e. each potential cluster head node sends a message to the edge node in order to evaluate the link quality before doing the selection. 
Therefore, those works use a \textit{locally-centralized} strategy for resource discovery. 

However, using a locally-centralized strategy comes at the cost of the necessity to regularly update the node gathering the resource information. Such updates are costly, for example in terms of energy consumption, as studied by Liu et al. \cite{AdaptiveResourceDiscovery}. They propose an algorithm enabling a switch between a locally-centralized mode and a \textit{distributed} mode. In the locally-centralized mode, end devices propagate their resource information/request to a specific node. In the distributed mode, end devices look for resources in the neighboring devices by using ad-hoc WLAN. They qualify their strategy as adaptive as it takes into account the current characteristics of resource distribution and usage in the network. When evaluating the energy consumption of two variants of the adaptive strategy, these perform close to the ideal energy consumption (10\% to 13\% more energy) and both perform better  
than strategies using only a distributed or locally-centralized mode.  

Finally, Zamani et al. \cite{DeadlineConstrainedVideoAnalysis} use a framework called CometCloud which performs resource discovery for video analysis and compare the benefit gained to a solution in the cloud. 

\subsection{Resource allocation}

\begin{figure}[!t]
\centering
\newcommand{\slice}[4]{
  \pgfmathparse{0.5*#1+0.5*#2}
  \let\midangle\pgfmathresult

  \draw[thick,fill=black!10] (0,0) -- (#1:1) arc (#1:#2:1) -- cycle;

  \node[label=\midangle:#4] at (\midangle:1) {};

  \pgfmathparse{min((#2-#1-10)/110*(-0.3),0)}
  \let\temp\pgfmathresult
  \pgfmathparse{max(\temp,-0.5) + 0.8}
  \let\innerpos\pgfmathresult
  \node at (\midangle:\innerpos) {#3};
}

\begin{tikzpicture}[scale=2]

\newcounter{a}
\newcounter{b}
\foreach \p/\t in {14/Placement, 7/Migration, 3/Scheduling,
                   6/Multi-perspective}
  {
    \setcounter{a}{\value{b}}
    \addtocounter{b}{\p}
    \slice{\thea/30*360}
          {\theb/30*360}
          {\p}{\t}
  }

\end{tikzpicture}
\caption{Distribution of resource allocation approaches in the surveyed articles.}
\label{fig_allocation}
\end{figure}
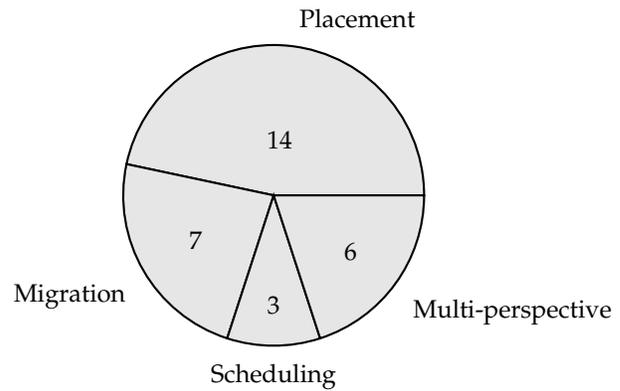

Resource allocation can be tackled from two different perspectives: \emph{where} to allocate (both initially, but also where and when to perform a migration if needed), and \emph{when and how much} to allocate.
Among the dominant approaches to allocation, we find the following three perspectives: placement (14 articles), migration (7 articles), scheduling (3 articles), as well as a multi-perspective one (6 articles) as shown in Figure \ref{fig_allocation}.

In what follows we group papers that have a single perspective under subsections \ref{sec:Placement}, \ref{sec:Migration} and \ref{sec:Scheduling}, and then move on to papers where several perspectives are present.

\subsubsection{Placement}
\label{sec:Placement}

Most of the surveyed works emphasize the first perspective, i.e. where should the task be executed and the resource allocated for the best possible execution. The definition of best execution varies depending on the considered system and the focus of the research.

Load distribution to achieve lower latency has attracted attention in a number of surveyed works, and it can be seen as an instance of placement. Fricker et al. \cite{OffloadingSchemeDCFog} propose an offloading strategy between edge data centers under high loads which show the benefit of having a larger data center as back-up for a small one. Latency is also the focus of study for Borylo et al. \cite{EnergyAwareFogAndCloud}  who investigate dynamic resource provisioning. They present a policy in which the edge can use the cloud in compliance with the latency requirements of the edge but enables a better energy efficiency by using resources in data centers powered by green energy.

Also focusing on energy, Mtibaa et al. \cite{TowardsResourceSharing} propose a power balancing algorithm in which a device decides whether to offload and to which other device depending on the energy left in the devices' batteries. In a single hop scenario, their solution extended the time before the first device of the group runs out of battery by 60\% (from 40 minutes to 2 hours) compared to a greedy solution.  

Oueis \cite{OueisThesis} tackles the issue of load distribution and resource allocation in small cell clusters. She formulates a joint computational and communication resource allocation and optimization problem in a multi-user case with a focus on latency and power efficiency. Similarly, Sardellitti et al. \cite{Sardellitti_JointOptimization} study an offloading problem when the end users are separated into two groups: those who need computation offloading and those who don't. They propose a method to jointly optimize communication and computation resources are both user groups compete for communication resources but only the first group compete for computation resources. They first present an algorithm for the single-user case and then two algorithms for the multiple cells case, a centralized one and a distributed version to mitigate the communication overhead induced by the centralized approach. 

Valancius et al. \cite{GreeningNanoDataCenters} propose a content placement strategy where the content is movies. The focus is first on finding the optimal number of replicas of the data to be stored, and then on placing the replicas on available gateways. Similarly, Qi et al. \cite{DynamicResourceOrchestration} present an allocation scheme where the resource (either coming from a cloudlet or a cloud) is chosen for each task. The aim is to pick the resource from the most suitable location when the user is moving. 

Wang et al. \cite{Wang_DynamicServicePlacementJournal} study service placement in a system composed of edge server nodes and traditional cloud nodes. Simulation results with real-world traces from San Francisco taxis show that the proposed approach is close to the case of online placement when the future is known, outperforming edge-only or cloud-only solutions. Similarly, Skarlat et al. \cite{TowardsQoSAwareFogServicePlacement} present a service placement problem for IoT services.

Mascitti et al. \cite{ServiceProvisioningOpportunistic} present an algorithm where an end device can choose to either use a resource from a node it is directly in contact with or to compose different resources from different nodes in order to complete its task. Coming from the same research group, Borgia et al. \cite{MobileEdgeCloudsForInformationCentric} present a framework where the decision is taken to obtain a service from a local group of end devices or from the cloud, based on an estimation of the time required to obtain the service.  

Confais et al. \cite{ObjectStore} propose a storage mechanism where the objects to be stored are primarily placed locally at their creation but can then be copied to another location if another edge site is requiring access to it. In vehicular networks where resources are mobile, placement has to take account of changes in location. Meng et al. \cite{ResourecAllocSchemeVehicular} present a resource allocation scheme which can manage resources from a vehicular cloud, the edge, or the cloud. Their focus is on minimizing delays for the users. 

In the application domain of healthcare, Gu et al. \cite{CostEfficient} include VM placement in their optimization problem for analyzing data. Their two-phase solution has a nearly optimal solution  and outperforms a greedy strategy with regards to cost.

\subsubsection{Migration}
\label{sec:Migration}

Still considering where the task should be executed, when it comes to virtual entities such as services, applications, tasks, and VMs, the focus could also be on how they can be moved during execution if the new location is better, i.e. on migration. 

For example, Tärneberg et al. \cite{DynamicApplicationPlacement} study application-driven placement and present a system model for mobile cloud network with a dynamic placement algorithm that guarantees application performance, minimizes cost, and tackles resource asymmetry problems.
Plachy et al. \cite{DynamicResourceAlllocation} propose a cooperative and dynamic VM placement algorithm associated with another cooperative algorithm for selecting a suitable communication path. They use VM migration to solve user mobility problems.  

Other works focus specifically on the problem of migrating resources. For example, Gomes et al. \cite{EdgeCachingMobilityPrediction} present a content-relocation algorithm for migrating the content of caches present in edge devices. This needs a prediction of user mobility.  

With respect to virtual computational resources, Fan et al. \cite{EnergyDrivenAvatar} and Rodrigues et al. \cite{HybridMethodServiceDelay} focus on VM migration, but with different optimization objectives (increase the use of green energy and minimize delays, respectively).
Yousaf et al. \cite{FinedGrainedResourceAware} propose a VM migration (and VM management) system that takes into account the relationship between resource units when making migration decisions. They present this work in the context of 5G but it should be applicable to all physical machines hosting VMs.

Finally, Penner et al. \cite{TransientCloud} introduce the term \emph{transient cloud} and concentrate on task assignment towards a given node. 
They present a collaborative computing platform that allows nearby devices to form an ad-hoc network and provide the ability to balance the assignments among themselves. This may be considered as a form of migration.

\subsubsection{Scheduling}
\label{sec:Scheduling}

While there is a huge body of research available on when and how many resources to allocate within networking and cloud-specific areas, our goal was here to identify examples where scheduling decisions are at the edge level in the sense of our terminology in section ~\ref{sec:sub_terminology}.

Regarding when to allocate resources, Bittencourt et al. \cite{MobilityAwareApplicationScheduling} study the impact of three different fog scheduling strategies on application QoS (Concurrent, First Come-First Served, and Delay-priority). For two applications studied, when more than four users are moved between cloudlets, the concurrent strategy exhibits a lot longer delays than the sequential ones. However, this same strategy is using the network a lot less than the other two. 

With the same focus, Singh et al. \cite{Singh_RTSANE} consider only scheduling for tasks with a \textit{private} tag. Those can only be executed on the local edge server and will be rejected if not enough resources are available. In their algorithm, tasks are considered in an earliest-deadline-first manner.

Regarding how many resources to allocate, Wang et al. \cite{CostEffectiveResourceAllocaion} propose a joint cost-effective resource allocation between the mobile cloud computing infrastructures and the cloud radio access network infrastructure. If the need of the application is greater than the available computational resources, then they reduce the amount given to each virtual machine so that it fits the total amount available and adapt the data rate accordingly. They show that joint optimization with respect to cost and energy performs better compared to separate cost- and energy-optimization strategies.

Similarly, Wang et al. \cite{Wang_ElasticUrbanVideosur} propose elastic resource allocation for video-surveillance systems. The elasticity comes from an algorithm they propose to handle some emergency surveillance event (like tracking a criminal) which requires a sudden increase of computation and communication resources to make sure that all the possible images are analyzed within a reasonable timeframe. When such an emergency event happens, network bandwidth allocation is reconfigured and computing resources are reallocated (by launching new VMs in the impacted zone and balancing the workload on nodes). When experimenting in their physical testbed, they verified that data propagation round-trip time is about 5 times lower with edge nodes close to the cameras compared to the cloud. They also found that the time for launching new VMs in the emergency mode is between one and two minutes, which they claim is acceptable in such a scenario.

When addressing scheduling, the surveyed articles most often do it at the same time as placement or migration, which is the topic of the next subsection.

\subsubsection{Multiple perspectives}

A work that tackles both perspectives is by Liu et al. \cite{AdaptiveMultiResourceAllocationCloudlet}. It presents a multi-resource allocation system which first decides whether the request should be served or rejected (admission control), then where to run it (edge or cloud level), and finally how much bandwidth and computing resources should be allocated for this task. To do that, they use Semi-Markov Decision Processes and their aim is to maximize system benefit while guaranteeing QoS for the users. To measure the benefit, they use blocking probability and service time as metrics. When evaluating, they compare to two greedy strategies and show that their proposal outperforms the first one and provides a 90\% reduction of the blocking probability with only a slight increase of service time compared to the second greedy strategy, which would be acceptable for congested situations. 

In the context of video analysis, Zamani et al. \cite{DeadlineConstrainedVideoAnalysis} also studied those two perspectives. Their scheduling is based on identified chunks of video, applying two alternatives: minimizing computation time or minimizing computation costs. Their placement is done after resource discovery using CometCloud. In their evaluation, they showed that the solution using edge accepts more tasks and in particular more high-value tasks than a solution using only the cloud. Hence, the overall value obtained from the processed data is maximized at the same time as the throughput of the infrastructure. 

Also in the area of video analytics, Yi et al. \cite{Yi_LAVEA} investigate three task prioritizing schemes for scheduling the task requests at a receiving edge node. Their solution,  using the flow job shop model and applying a well-known approach (Johnsson's rule), aims at minimizing the makespan. Their simulations compared the approach with other strategies (Short IO First, Longest CPU Last) and found that response time was improved.  Their work also includes a second perspective, by investigating three task placement schemes for collaboration within the edge level (Shortest Transmission Time first, Shortest Queue Length First, and Shortest Scheduling Latency First). Using their testbed, they found that the Shortest Scheduling Latency First achieves the best performance in terms of task completion time.

Singh et al. \cite{Singh_RTSANE} consider both placement and scheduling with respect to semi-private or public tasks (in addition to what was mentioned in the last subsection for private tasks). Those tasks are placed after a decision is taken for the private ones. Still considering earliest-deadline first, the placement strategy is to try first one's own edge, then one's own cloud and if they are overloaded, go to some external edge, and then to an external cloud. In the evaluation, they show that for tasks having tight deadlines, their system RT-SANE will complete a lot more tasks before their deadline than a cloud-only solution.

How many resources are to be allocated to a given IoT data generator is a topic of discussion by Arkian et al. \cite{MIST}, in which they first mathematically model deployment and communication costs on various fog nodes and then decide on placement of VMs to achieve lowest costs. Analyzing monetary costs for compute nodes, their fog solution decreased the cost by over 33\% compared to using a cloud solution. For routing and storage monetary cost, the decrease is about 20\%. 

Habak et al. \cite{Habak_WorkloadManagement} first consider  placement for deciding in which end device a task will be run. They use a path-based assignment policy with the aim of minimizing the overhead of transmitting data needed for task execution between end devices. In the evaluation, this translates into performing better than two other baseline solutions in terms of service completion time. Then, they also consider scheduling of the computation resource. This should be done in a predictable way so that the part of the system distributing the tasks can make good decisions. They propose a fair queuing based task pick-up that ensures  a fair execution of the tasks belonging to different services. Moreover, they implement an early pick-up mechanism to enhance the previous mechanism so that a task with an urgent deadline but belonging to a service with a lower priority can execute before a higher priority task if this one still meets its deadline.

While this is not the focus of the survey, and as such is not included in the table, Dong et al. \cite{Dong_EnergyEfficient} study offloading and Earliest Deadline First scheduling within end devices. They find that one of their proposed approaches maintains good predictability for twice higher CPU utilization than widely-used approaches, while keeping energy consumption reasonable.

\subsection{Resource sharing}

Resources on end devices are heterogeneous and most of the time scarce, and edge devices also have limited resources compared to (almost infinite) resources in the cloud. Sharing resources between devices or between end and edge devices aims at tackling three different issues: not having the needed resource at all in the device where the task is initiated, not having enough of it, or using other devices' resources in order to get a faster completion of the task.

Sharing resources is typically realized by pooling resources in the local vicinity of client nodes. This can extend to the edge domain (clustering edge servers) or remain at end devices. The latter is investigated by Skarlat et al in so-called \emph{fog colonies} \cite{TowardsQoSAwareFogServicePlacement}, by Arkian et al. within vehicular clusters \cite{ClusterBasedVehicularClouds}, or by Bianzino et al. ~\cite{Bianzino2014} for uploading data streams in presence of mobility.

We can classify the surveyed articles into two categories according to whether they include how to form the groups of devices that will share resources or if they assume that the formation is already done and focus on the actual sharing. We call these two categories as \textit{dynamic coalitions}, and \textit{static coalitions} 
respectively.

Starting with dynamic coalitions, Chen et al. \cite{Chen_sociallyTrusted} and Bianzino et al. \cite{Bianzino2014} include the formation of device coalitions. Chen et al. \cite{Chen_sociallyTrusted} do it using a coalition game incorporating trust considerations. When supply matches demand, they found that using a coalition can lead up to 40\% lower weighted cost (including latency and monetary considerations) compared to a non-cooperative scenario. When there is overload or light workload, it is either not possible or not needed to collaborate and the gain is very low. Bianzino et al \cite{Bianzino2014} express resource sharing as an optimization problem where the aim is to create as few and large groups as possible to minimize the number of high-energy interfaces that will be used. They evaluate that their algorithm leads to over 60\% energy saving of the total energy consumed by the end devices.  

Still using dynamic coalitions, Arkian et al. \cite{ClusterBasedVehicularClouds} and Athwani et al. \cite{ResourceDiscoveryInMCC} propose methods to create clusters. The former compare their method to an earlier baseline and achieve  3 times lower service discovery delay and 4,5 times lower service consumption delay for a small number (50) of vehicles. The latter show that energy consumption is similar to a centralized approach while the delay is closer to a flooding approach (i.e. low in both cases).    

However, creating and maintaining a group of devices which can share their resources has a cost, for example shown by Athwani et al.  \cite{ResourceDiscoveryInMCC} who concluded that maintaining the cluster consumes extra energy, especially if the devices are very mobile. This is why it is beneficial to do the resource sharing in two phases, where the first phase is deciding whether the device gains more by working alone or joining a coalition, and the second one is deciding if the device will consume others' resources \cite{Bianzino2014,CooperativeResourceManagementVehicular}. Yu et al. \cite{CooperativeResourceManagementVehicular} show that their cooperative solution improves user QoS (defined by how much computing and bandwidth resources are allocated to a user) by 75\%. However, this paper is using traditional cloud resources and not edge so it is not included in the tables.

Moving to static coalitions, Skarlat et al. \cite{TowardsQoSAwareFogServicePlacement} consider resources shared between two neighbor fog colonies and achieve a 35\% reduction of execution cost compared to a cloud-only strategy. With regard to data, a resource that many stakeholders may be interested in sharing, Zhang et al. \cite{WeisongFirework} present a framework for this type of sharing, called Firework. They include two case studies, including the search for a person with the help of multiple cameras from different owners. 

Some researchers, such as Liu et al.  \cite{OpportunisticResourceSharingMCC}, try to exploit opportunistic contacts between the devices, creating a resource sharing mechanism that enables faster task completion.
They propose different models for calculating task latencies and their approximation algorithm performs better than two other strategies. Similarly, Mtibaa et al. \cite{TowardsResourceSharing} define three mobile device clusters (one hop, two hop and opportunistic) which can share their resources. Their aim is to share resources in order to get the longest possible network lifetime, i.e. saving as much energy as possible through offloading to another device so that the devices can stay on longer. They identify two important topological factors: number of hops and disconnection rate due to mobility.

Resource sharing can perhaps speed up the execution of a task, but Nishio et al. \cite{ServiceOrientedHeterogeneousResourceSharing} argue that this is not bringing any advantage for the user if we do not consider task dependencies in order to provide a service to the user. They provide the example of a GPS service: if the best route calculation is very fast but the downloading of the map is not, the service to the user won't get faster as both are needed. 
Habak et al. \cite{Habak_WorkloadManagement} consider sharing of end device resources belonging to a femtocloud in order to execute tasks. In their system, the owner of the end device can configure how they want to share resources via their personalized resource sharing policies.

Finally, even if resource sharing can bring benefits for a group of end devices, it is not obvious that users will agree to share their resources, especially if they are always on the providing side. Therefore there is a need to develop incentives for resource sharing such as works by Tang et al. \cite{DoubleSidedBidding}, Bianzino et al. \cite{Bianzino2014}, and Chen et al.\cite{Chen_sociallyTrusted}.
The following mechanisms are provided in the above works respectively: 
\begin{enumerate}[label=\alph*)]
\item a double bidding mechanism for demander and supplier of resources where the focus is on how to encourage mobiles with resources to share them.
\item a mechanism for lending energy to vicinity nodes is rewarded and can be used in future scenarios when the lending node itself needs energy.
\item payment incentives for lending out resources.
\end{enumerate}

On the same topic, Habak et al. \cite{Habak_WorkloadManagement} performed a pilot study to identify effective incentive mechanisms. They studied the willingness of around 50 students to share their resources in 4 scenarios and found out that they would agree to share their resources if they are getting compensation (for example money) for it or if the reason for the computation taking place is significant (for example emergencies).  

\begin{table*}
\centering
\caption{Surveyed articles according to resource type and objective of resource management.} 
\label{table_type_objective}
\begin{tabular}{@{} |c|l|p{2cm}|p{2cm}|p{2cm}|p{2cm}|p{2cm}| @{}} 
\cline{3-7}
\multicolumn{1}{c}{} & \multicolumn{1}{c|}{}& \multicolumn{5}{c|}{Objective} \\ 
 \cline{3-7} 
\multicolumn{1}{c}{} & \multicolumn{1}{c|}{} & \parbox{1cm}{\centering Resource\\estimation} & \parbox{1cm}{\centering Resource\\discovery} & \parbox{1cm}{\centering Resource\\allocation} & \parbox{1cm}{\centering Resource\\sharing} & \parbox{1cm}{\centering Resource\\optimization} \\ 
\hline
\multirow {6} {*} {\rot{\parbox{8cm}{\centering Resource type}}}& 
Computation     &
\cite{Habak_WorkloadManagement}&
\cite{DeadlineConstrainedVideoAnalysis, ResourceDiscoveryInMCC, ClusterBasedVehicularClouds}&
{\cite{AdaptiveMultiResourceAllocationCloudlet, MIST, EnergyDrivenAvatar, OueisThesis, FinedGrainedResourceAware, CostEffectiveResourceAllocaion, CostEfficient, DynamicApplicationPlacement, DynamicResourceAlllocation, OffloadingSchemeDCFog, HybridMethodServiceDelay, MobilityAwareApplicationScheduling, DeadlineConstrainedVideoAnalysis, Wang_ElasticUrbanVideosur, Yi_LAVEA, Sardellitti_JointOptimization, Singh_RTSANE, TowardsQoSAwareFogServicePlacement, Habak_WorkloadManagement, ServiceProvisioningOpportunistic, ResourecAllocSchemeVehicular}}
&
\cite{DoubleSidedBidding, WeisongFirework, Chen_sociallyTrusted, ServiceOrientedHeterogeneousResourceSharing, TowardsQoSAwareFogServicePlacement, ClusterBasedVehicularClouds, Habak_WorkloadManagement, ResourceDiscoveryInMCC} &
\cite{AdaptiveMultiResourceAllocationCloudlet, MIST, EnergyDrivenAvatar, OueisThesis, FinedGrainedResourceAware, CostEffectiveResourceAllocaion, CostEfficient, DynamicApplicationPlacement, HybridMethodServiceDelay, DeadlineConstrainedVideoAnalysis, Yi_LAVEA, Sardellitti_JointOptimization, ServiceOrientedHeterogeneousResourceSharing, TowardsQoSAwareFogServicePlacement, ResourceDiscoveryInMCC, ClusterBasedVehicularClouds, Habak_WorkloadManagement, ResourecAllocSchemeVehicular} \\
\cline{2-7} 
& Communication     &
\cite{Habak_WorkloadManagement}&
\cite{DeadlineConstrainedVideoAnalysis, ResourceDiscoveryInMCC, ClusterBasedVehicularClouds}&
\cite{AdaptiveMultiResourceAllocationCloudlet, ObjectStore, MIST, OueisThesis, FinedGrainedResourceAware, CostEffectiveResourceAllocaion, CostEfficient, DynamicApplicationPlacement, DynamicResourceAlllocation, HybridMethodServiceDelay, MobilityAwareApplicationScheduling, DeadlineConstrainedVideoAnalysis, GreeningNanoDataCenters, Wang_ElasticUrbanVideosur, Yi_LAVEA, Sardellitti_JointOptimization, Singh_RTSANE, TowardsQoSAwareFogServicePlacement, MobileEdgeCloudsForInformationCentric, Habak_WorkloadManagement, ServiceProvisioningOpportunistic, ResourecAllocSchemeVehicular}&
\cite{DoubleSidedBidding, Chen_sociallyTrusted, ServiceOrientedHeterogeneousResourceSharing, TowardsQoSAwareFogServicePlacement, ClusterBasedVehicularClouds, Bianzino2014, Habak_WorkloadManagement, OpportunisticResourceSharingMCC, ResourceDiscoveryInMCC} & 
\cite{AdaptiveMultiResourceAllocationCloudlet, MIST, OueisThesis, FinedGrainedResourceAware, CostEffectiveResourceAllocaion, CostEfficient, DynamicApplicationPlacement, HybridMethodServiceDelay, DeadlineConstrainedVideoAnalysis, GreeningNanoDataCenters, Yi_LAVEA, Sardellitti_JointOptimization, ServiceOrientedHeterogeneousResourceSharing, TowardsQoSAwareFogServicePlacement, ResourceDiscoveryInMCC, ClusterBasedVehicularClouds, Habak_WorkloadManagement, OpportunisticResourceSharingMCC, ResourecAllocSchemeVehicular,Bianzino2014} \\
\cline{2-7} 
& Storage    &
&
\cite{ClusterBasedVehicularClouds}&
\cite{ObjectStore, MIST, FinedGrainedResourceAware, CostEfficient, GreeningNanoDataCenters, Wang_ElasticUrbanVideosur, TowardsQoSAwareFogServicePlacement}&
\cite{TowardsQoSAwareFogServicePlacement, ClusterBasedVehicularClouds}& 
\cite{MIST, FinedGrainedResourceAware, CostEfficient, GreeningNanoDataCenters, TowardsQoSAwareFogServicePlacement, ClusterBasedVehicularClouds} \\
\cline{2-7} 
& Data    &
\cite{Habak_WorkloadManagement}&
&
\cite{EdgeCachingMobilityPrediction, TowardsQoSAwareFogServicePlacement, MobileEdgeCloudsForInformationCentric, Habak_WorkloadManagement}&
\cite{WeisongFirework, ServiceOrientedHeterogeneousResourceSharing, TowardsQoSAwareFogServicePlacement, Habak_WorkloadManagement}& 
\cite{ServiceOrientedHeterogeneousResourceSharing, TowardsQoSAwareFogServicePlacement, Habak_WorkloadManagement} \\
\cline{2-7} 
& Energy    &
\cite{AdaptiveResourceDiscovery, TowardsResourceSharing}&
\cite{ResourceDiscoveryInMCC, AdaptiveResourceDiscovery}&
\cite{EnergyDrivenAvatar, OueisThesis, EnergyAwareFogAndCloud, GreeningNanoDataCenters, Sardellitti_JointOptimization, DynamicResourceOrchestration, TowardsResourceSharing}&
\cite{Chen_sociallyTrusted, ServiceOrientedHeterogeneousResourceSharing, Bianzino2014, TowardsResourceSharing, ResourceDiscoveryInMCC}& 
\cite{EnergyDrivenAvatar, OueisThesis, GreeningNanoDataCenters, Sardellitti_JointOptimization, ServiceOrientedHeterogeneousResourceSharing, ResourceDiscoveryInMCC, AdaptiveResourceDiscovery, DynamicResourceOrchestration, TowardsResourceSharing,Bianzino2014} \\
\cline{2-7} 
& Generic     &
\cite{PREFog, FogDynamicResourceEstimation, Wang_DynamicServicePlacementJournal, AdaptiveResourceDiscovery}&
\cite{AdaptiveResourceDiscovery}&
\cite{Wang_DynamicServicePlacementJournal, MobileEdgeCloudsForInformationCentric, TransientCloud, DynamicResourceOrchestration}&
\cite{OpportunisticResourceSharingMCC}& 
\cite{Wang_DynamicServicePlacementJournal, AdaptiveResourceDiscovery, OpportunisticResourceSharingMCC, DynamicResourceOrchestration}\\
\hline
\end{tabular}
\end{table*}
 
\subsection{Resource optimization}

A fifth objective pursued in the surveyed works is to optimize the resource use at the edge. This is usually a joint objective together with one of the previously described objectives. 
Which aspect should be optimized and the associated constraints varies among the surveyed works but the three main ones are QoS (often understood as latency), energy, and operational cost. 
How the optimization problem is formulated and solved also varies, and we present those variations in this section. 

First, some articles consider selecting the optimum solution by comparing the results from different candidates and selecting the minimum/maximum value depending on the objective. For example, Yousaf et al. \cite{FinedGrainedResourceAware} select the value maximizing the resource utilization, Athwani et al. \cite{ResourceDiscoveryInMCC} use the minimum value of a custom function to select the cluster head, and Mtibaa et al. \cite{TowardsResourceSharing} select the configuration maximizing the estimated remaining energy. 

Another group of works solves their optimization problem using linear programming \cite{AdaptiveMultiResourceAllocationCloudlet,GreeningNanoDataCenters} or an approximation based on linear programming \cite{OpportunisticResourceSharingMCC}.

A third group of works uses integer linear programming \cite{TowardsQoSAwareFogServicePlacement,Bianzino2014} or
mixed-integer linear programming \cite{EnergyDrivenAvatar}. 
Qi et al. \cite{DynamicResourceOrchestration} formulate their task allocation problem using integer programming and solve it by a self-adaptive learning particle swarm optimization algorithm. 
First formulating using mixed-integer non-linear programming, Arkian et al. \cite{MIST} then linearize the problem and solve it using mixed-integer linear programming. Gu et al. \cite{CostEfficient} do the same and then use heuristics.
Using a different approach, Yi et al. \cite{Yi_LAVEA} first formulate a mixed integer non-linear programming problem but then relax the integer constraints and use sequential quadratic programming for solving. 

Some works focus on convex problems, like Wang et al.\cite{Wang_DynamicServicePlacementJournal} who use an approximation algorithm in the online case and Nishio et al. \cite{ServiceOrientedHeterogeneousResourceSharing} who use a heuristic. 
Starting with non-convex problems, Oueis \cite{OueisThesis} cast them into convex ones and  
Wang et al. \cite{CostEffectiveResourceAllocaion} first use a Weighted Minimum Mean Square Error-based method on their non-convex problem to obtain a convex problem that they apply the block coordinate descent method to for solving. 
Finally, Sardellitti et al. \cite{Sardellitti_JointOptimization} have an optimization problem in the multiple-cells case that is non-convex and they solve it by developing a method based on Successive Convex Approximation for the centralized approach. For the distributed approach, they choose the approximation functions in a way that allows decomposition in smaller subproblems solvable in parallel.

A further group of works proposes their own algorithm or heuristic. Tärneberg et al. \cite{DynamicApplicationPlacement} approximate an exhaustive search approach yielding an optimal solution but having exponential computation complexity with an iterative local search algorithm finding a local optimal solution. Zamani et al. \cite{DeadlineConstrainedVideoAnalysis} implement an optimization strategy where constraints on computation time and cost are enforced using an admission control strategy. Wang et al. \cite{Wang_DynamicServicePlacementJournal} present a  binary search algorithm for finding the optimal look-ahead window size, and Habak et al. \cite{Habak_WorkloadManagement} propose an algorithm in order to do deadline-based optimization when a helper has to handle multiple tasks belonging to different services. Finally, Liu et al. \cite{AdaptiveResourceDiscovery} propose a heuristic algorithm that uses different statistics to estimate the energy that is going to be consumed in each of the two possible modes during a time slot, and chooses which mode to use depending on this and other parameters.

Other methods can be used to compare heuristics with baselines, or to solve a formulation in a custom form. In the offline case, Wang et al. \cite{Wang_DynamicServicePlacementJournal} show that their problem is equivalent to the shortest-path problem and solve it by using dynamic programming. Meng et al. \cite{ResourecAllocSchemeVehicular} solve Bellman equations recursively, Rodrigues et al. \cite{HybridMethodServiceDelay} use integration techniques, and Arkian et al. \cite{ClusterBasedVehicularClouds} consider fuzzy logic and Q-learning. 

\subsection{Summary of objectives in resource management}
\label{sec:SummaryObjective}

By far, the most active area of research in the edge resource management is resource allocation, as visible in Table~\ref{table_articles_objective}. This is followed by optimization as a goal, where we see a great majority of papers present. Among the objectives from our taxonomy, resource estimation and resource discovery are least studied. Resource sharing, to the extent it is used, is well-represented among the second and third type of architectures in Figure~\ref{fig_architectures}, i.e. coordinator device, and device clouds, but not in the first type of architecture (edge server). 

Somewhat surprisingly, while scheduling is a major topic in cloud systems, the edge-specific literature does not consider it as the main problem, as evident from fewer works addressing scheduling compared to placement and migration. Where autoscaling is mentioned in an edge context, authors typically deal with offloading to the cloud which was not the focus of our work. There are several excellent surveys already covering these. The work by Wang et al.~\cite{ENORM}, addressing autoscaling and the edge is among few exceptions, so we did not create a special category for this type of work.

While the previous breakdown was done in a resource independent manner, it is also interesting to consider the resource type studied with regards to the resource management objectives. Table \ref{table_type_objective} thus combines the information contained in Tables \ref{table_articles_types} and \ref{table_articles_objective} to give us this view. 
Not surprisingly, most of the articles consider computation and communication for resource allocation and optimization. Quite expected as well, the proportion of resource sharing articles (from Table \ref{table_articles_objective}) considering energy as a resource (45\% according to Table \ref{table_type_objective}) is higher than the proportion of, for example, resource allocation articles considering energy (23\%), as an incentive to share resources is when you consider energy-constrained devices. It is interesting to note that in the surveyed works, resource estimation is most often done for a generic resource type and that none of the articles combined resource estimation and storage, and resource discovery and data.

\section{Resource location}
\label{sec:ResourceLocation}

Computing at the edge differentiates itself from regular cloud computing with the fact that resources used can belong to different levels. It is indeed not uncommon to use resources at the edge level primarily, but also from the cloud level if required. Moreover, end devices, and sometimes edge devices do not have to be stationary as in a data center. Note that here we make a distinction between mobility on the demand side and mobility on the supply side. Even though the demand side clients are almost always mobile, the infrastructure that supplies the adequate resources has been invariably stationary in the past. 

In this section, we first look at where the \textit{managed} resources considered are located within the architectures presented in Figure \ref{fig_architectures}. We then shift focus and look at the same set of resources again but this time studying their mobility. 
 
\begin{table*}
\centering
\caption{Managed resources and their supply-side mobility.} 
\label{table_articles_resourcelocation}
\begin{tabular}{@{} cl|c|c|c| @{}} 
\cline{3-5}
 & & \multicolumn{3}{c|}{Managed resources' location} \\ 
  \cline{3-5}
 & Article & Device level & Edge level & Cloud level\\ 
\hline
\multirow {23} {*} {\rot{Edge server}}  & Liu \cite{AdaptiveMultiResourceAllocationCloudlet} &&Stationary&Stationary \\
& Confais \cite{ObjectStore}      &&Stationary&Stationary \\
& Aazam \cite{PREFog}      &&Stationary& \\
& Arkian \cite{MIST}      &&Stationary& \\
& Aazam \cite{FogDynamicResourceEstimation}    &&Stationary+Mobile& \\
& Fan \cite{EnergyDrivenAvatar}      &&Stationary& \\
& Oueis \cite{OueisThesis}      &&Stationary&\\
& Tang \cite{DoubleSidedBidding}      &Stationary&&\\
& Borylo \cite{EnergyAwareFogAndCloud}      &&Stationary&Stationary \\
& Yousaf \cite{FinedGrainedResourceAware}      &&Stationary& \\
& Wang \cite{CostEffectiveResourceAllocaion}      &&Stationary &\\
& Gu \cite{CostEfficient}      &&Stationary& \\
& Tärneberg \cite{DynamicApplicationPlacement}      &&Stationary &\\
& Plachy \cite{DynamicResourceAlllocation}      &&Stationary& \\
& Gomes \cite{EdgeCachingMobilityPrediction}      &&Stationary& \\
& Fricker \cite{OffloadingSchemeDCFog}      &&Stationary& \\
& Rodrigues \cite{HybridMethodServiceDelay}      &&Stationary& \\
& Zhang \cite{WeisongFirework}      &Stationary&Stationary&\\
& Bittencourt \cite{MobilityAwareApplicationScheduling}      &&Stationary&Stationary \\
& Zamani \cite{DeadlineConstrainedVideoAnalysis}      &Stationary&Stationary&Stationary \\
& Valancius \cite{GreeningNanoDataCenters}   &&Stationary&Stationary   \\
& Chen \cite{Chen_sociallyTrusted} &&Stationary&   \\
& Wang \cite{Wang_ElasticUrbanVideosur} &&Stationary&   \\
& Yi \cite{Yi_LAVEA} &&Stationary& Stationary  \\
& Wang \cite{Wang_DynamicServicePlacementJournal} &&Stationary& Stationary  \\
& Sardellitti \cite{Sardellitti_JointOptimization} &&Stationary&   \\
& Singh \cite{Singh_RTSANE} &&Stationary&Stationary   \\
\hline
\multirow {8} {*} {\rot{\parbox{1.5cm}{\centering Coordinator \\ device}}}   & Nishio \cite{ServiceOrientedHeterogeneousResourceSharing}   &Stationary&&\\ 
& Skarlat \cite{TowardsQoSAwareFogServicePlacement}      &&Stationary&Stationary \\
&Borgia \cite{MobileEdgeCloudsForInformationCentric}     &&Mobile&Stationary \\ 
& Athwani \cite{ResourceDiscoveryInMCC}      &&Mobile& \\
& Arkian \cite{ClusterBasedVehicularClouds}      &&Mobile& \\
& Penner \cite{TransientCloud}      &&Mobile&\\
& Bianzino \cite{Bianzino2014} &Mobile&Mobile&\\ 
& Habak \cite{Habak_WorkloadManagement} &Mobile&Mobile&   \\
\hline
\multirow {6} {*} {\rot{Device cloud}} & Liu \cite{AdaptiveResourceDiscovery} &\multicolumn{2}{c|}{Stationary}& Stationary \\
& Mascitti \cite{ServiceProvisioningOpportunistic} &\multicolumn{2}{c|}{Mobile}& \\
& Liu \cite{OpportunisticResourceSharingMCC}    &\multicolumn{2}{c|}{Mobile}& \\
& Meng \cite{ResourecAllocSchemeVehicular}    &\multicolumn{2}{c|}{Stationary+Mobile}& Stationary \\
& Qi \cite{DynamicResourceOrchestration}      &\multicolumn{2}{c|}{Mobile}&Stationary \\
& Mtibaa \cite{TowardsResourceSharing}      &\multicolumn{2}{c|}{Mobile}& \\
\hline
\end{tabular}
\end{table*}

\subsection{Location within the architecture}
Edge resource management is actually not only about managing resources located at the edge level, as a study of the managed resources' location in the surveyed work reveals. This study is presented in Table \ref{table_articles_resourcelocation}.
\subsubsection{Single-level}
As expected when surveying edge resource management papers, a large part (54\%) of those consider managed resources located only at the edge level, for example, the works by Arkian et al. \cite{MIST}, Fan et al.\cite{EnergyDrivenAvatar}, Gomes et al. \cite{EdgeCachingMobilityPrediction}, Yousaf et al. \cite{FinedGrainedResourceAware}, Chen et al. \cite{Chen_sociallyTrusted}, Sardellitti et al. \cite{Sardellitti_JointOptimization}, and Wang et al. \cite{Wang_ElasticUrbanVideosur}. 

Aazam et al. \cite{PREFog} consider resources located at only one physical location, a fog node, but considering resources within the same architectural level does most often not mean that the resources are located at the same physical location.
For example, Oueis \cite{OueisThesis} considers resources on different cells,  Gu et al. \cite{CostEfficient} and Plachy et al. \cite{DynamicResourceAlllocation} on different base stations. Fricker et al. \cite{OffloadingSchemeDCFog} and Rodrigues et al.  \cite{HybridMethodServiceDelay} consider task placement and migration on different types of edge devices (datacenters for the former and cloudlets for the latter).  

Essentially refining our architecture, some works distinguish different levels between the same architectural level from our Figure \ref{fig_architectures}. For example, Wang et al. \cite{CostEffectiveResourceAllocaion} consider transmission in the access network, and computation in a mobile cloud computing architecture. Tärneberg et al. \cite{DynamicApplicationPlacement} consider that data centers at the edge can have a different distance to the device, and different sizes.

Among the surveyed works, two works consider resources located only at the device level but where the management is performed at the edge: Tang et al. \cite{DoubleSidedBidding}, and Nishio et al. \cite{ServiceOrientedHeterogeneousResourceSharing} who consider resources present on different end devices. 

There is no work considering managed resources located at the cloud level only as those were on purpose considered out of the scope of this survey.

\subsubsection{Multi-level}

We observe that resources do not need to belong to the same architecture level.
Among the multi-levels works, the most common is to use resources located both at the edge and at the cloud level. This is the case in the works by Liu et al.  \cite{AdaptiveMultiResourceAllocationCloudlet}, Borylo et al. \cite{EnergyAwareFogAndCloud}, Valancius et al. \cite{GreeningNanoDataCenters}, Yi et al. \cite{Yi_LAVEA}, Wang et al. \cite{Wang_DynamicServicePlacementJournal}, and Singh et al. \cite{Singh_RTSANE}. Specifically, Skarlat et al. \cite{TowardsQoSAwareFogServicePlacement} and Bittencourt et al. \cite{MobilityAwareApplicationScheduling} favor using edge resources over cloud resources. Liu et al. \cite{AdaptiveResourceDiscovery} use resources in the device/edge level or in the cloud depending on the availability of the resources and Confais et al. \cite{ObjectStore} work with different storage locations at the edge or cloud level.

This is, however, not the only combination and Zhang et al. \cite{WeisongFirework} work with data as a resource with can be located both in the end devices and at the edge. This combination is also used by Bianzino et al. \cite{Bianzino2014} and Habak et al. \cite{Habak_WorkloadManagement} where an end device is promoted to an edge role.

Finally, combining the three levels, Zamani et al. \cite{DeadlineConstrainedVideoAnalysis} use resources on the device, on the network path to the cloud (edge level) and in the cloud level. 

\subsection{Resource mobility}
In an edge context, it is not obvious that resources located in the lower two levels of the architecture will be stationary or mobile. Therefore, it is interesting to study the mobility of the managed resources in the surveyed articles.

\subsubsection{Stationary resources}
Most of the surveyed articles (71\%) consider resources which are stationary only. This can be because the architecture/application considered does not have mobile resources or for simplification reasons. The latter is found in works where the architecture presented has resources that  are theoretically mobile but where this part is ignored in the solution or evaluation presented, e.g. in \cite{AdaptiveResourceDiscovery} or \cite{ServiceOrientedHeterogeneousResourceSharing}.

This preponderance of stationary resources may be explained by the fact that those works consider edge as an extension of the cloud, which has only stationary resources.

\subsubsection{Mobile resources}

Having mobile edge devices, and thus mobile resources obviously creates lots of challenges such as how to handle the unreliable connectivity of those resources, how to provide seamless handovers, etc. Thus, having mobile resources introduce another level of complexity in resource management algorithms.  

Different mobility models are used, for example, 
Penner et al. \cite{TransientCloud} model departure and arrival times using statistical models, which is similar to what is used by Bianzino et al. \cite{Bianzino2014}. Also using statistical models, Habak et al. \cite{Habak_WorkloadManagement} model arrival rate and presence time. In those statistical models, arrivals are modeled using a Poisson distribution, departure most often using an exponential distribution and presence time using a normal distribution. 
Another model that is relatively common is the Random Way Point Model, used by Mascitti et al.  \cite{ServiceProvisioningOpportunistic} and Liu et al. \cite{OpportunisticResourceSharingMCC}. 

With a different and more uncommon approach, Arkian et al. \cite{ClusterBasedVehicularClouds} consider the speed of the vehicles, and Athwani et al. \cite{ResourceDiscoveryInMCC} consider that 10\% of the nodes are moving after a request. 
Finally, Mtibaa et al. \cite{TowardsResourceSharing} consider both a mobility model with low disconnection rate and a mobility model based on a dataset (Infocom06) where the mobility of the devices is predictable in different communication scenarios.

\subsubsection{Combination of stationary and mobile resources}

Some works mention a combination of mobile and stationary resources. In the edge level, Aazam et al. \cite{FogDynamicResourceEstimation} consider different types of devices (stationary or mobile). However, the devices are actually not mobile in their simulations. 

Borgia et al. \cite{MobileEdgeCloudsForInformationCentric} consider the local cloud (i.e. the edge level) as mobile and the global cloud as stationary. They use the Random Way Point model for mobility. Similarly, Qi et al. \cite{DynamicResourceOrchestration} have mobile end devices and stationary infrastructure servers and describe their own mobility model. 
Meng et al. \cite{ResourecAllocSchemeVehicular} use a mobile vehicular cloud together with a stationary local cloud at the edge level and a stationary remote cloud. The mobility of the vehicles is modeled as a Poisson process.

\subsection{Summary of edge resource location}

Table \ref{table_articles_resourcelocation} reveals the distribution of the papers among the above categories, and clearly shows that fewer works are multi-level, and the biggest majority are stationary.
As noted before, less work are studying managed resources located at different levels and/or mobile. 

Note that this does not mean that the works do not consider mobility at all, only that the mobility is not on the supply side. Works including mobility on the demand side only are, for example, Plachy et al.  \cite{DynamicResourceAlllocation} who consider that computational resources needed by a user are allocated in a stationary base station in a VM, which can be transferred to another base station if the user is moving. Similar solutions are presented by Tärneberg et al. \cite{DynamicApplicationPlacement}, Gomes et al. \cite{EdgeCachingMobilityPrediction}, Oueis \cite{OueisThesis}, Fan et al. \cite{EnergyDrivenAvatar}, and Wang et al. \cite{Wang_DynamicServicePlacementJournal}. 

Despite demand side node mobility that may be present in all architectures, the supply side node mobility, i.e. the notion of mobile managed resource, is among the promises of what the edge brings. We see more mobile resources present in the second and third types of architecture (coordinator device and device cloud). It remains to be seen if the future works will include more 3-level works in which at least two are mobile.

\section{Resource use}
\label{sec:ResourceUse}
The final aspect of resource management considered in  our taxonomy is the purpose for which the resource will be used.  

\subsection{Functional properties}
Edge computing is promoted as a means of getting access to a given service in most of the surveyed articles, i.e. for satisfying functionality in an application.
There are numerous articles in the literature providing an overview of edge applications, including \cite{MaoSurveyPublished, Baktir_Survey,ChiangSurvey,FernandoSurvey,LiuSurvey, CloudletReferencePaper, FogReferencePaper, Satya2013}. Such applications range over augmented reality, 
connected vehicles, 
disaster recovery, 
and a lot of others.

When looking at the different applications used in the surveyed articles presented in the earlier sections, 
the first finding is that the majority of them (66\%) do not consider a specific application in their study. Instead, they refer to generic applications such as IoT services\cite{ObjectStore}, real-time applications\cite{DynamicResourceAlllocation}, latency-sensitive applications \cite{AdaptiveMultiResourceAllocationCloudlet}, or name some applications but only as an illustration.

Table \ref{table_applications} presents the remaining papers according to which type of application they consider. We can distinguish seven areas in which the described applications can be categorized. 
Note that in the Generic category we place papers that although not fixed towards one domain of application refer specifically to classes of applications that they exemplify clearly. 

\begin{table}
\centering
\caption{Applications considered in the surveyed articles.} 
\label{table_applications}
\begin{threeparttable}
\noindent
\begin{tabular}{@{} |p{1.9cm}|p{4cm}|p{1.75cm}| @{}} 
\hline
\textbf{Area}  & \textbf{Applications} & \textbf{Articles} \\ 
\hline
\multirow{2}{*}{Healthcare}  & Medical cyber-physical systems  & \cite{CostEfficient} \\ \cline{2-3}
& Connected health & \cite{WeisongFirework} \\ 
\hline
\multirow{3}{*}{Video}  & Video analytics  & \cite{DeadlineConstrainedVideoAnalysis,Yi_LAVEA} \\ \cline{2-3}

& Video surveillance & \cite{MobilityAwareApplicationScheduling,Wang_ElasticUrbanVideosur,WeisongFirework} \\ \cline{2-3}
& Video on Demand & \cite{GreeningNanoDataCenters} \\ 
\hline
\multirow{2}{*}{IoT}  & Crowd-sensing  & \cite{MIST} \\ \cline{2-3}
& Sense-Process-Actuate application & \cite{TowardsQoSAwareFogServicePlacement}\\
\hline
\multirow{1}{*}{Gaming}  & Electroencephalography
(EEG) tractor beam game  & \cite{MobilityAwareApplicationScheduling} \\
\hline
\multirow{1}{*}{Transportation}  & Connected vehicles  & \cite{ResourecAllocSchemeVehicular, ClusterBasedVehicularClouds} \\ \hline
\multirow{2}{*}{\parbox{2cm}{Content \\management}}  & User profiling  & \cite{EdgeCachingMobilityPrediction} \\
&&\\
\hline
\multirow{2}{*}{Generic}  & Computation/Communication intensive  & \cite{HybridMethodServiceDelay} \\ \cline{2-3}
 & Delay-sensitive/Delay-tolerant  & \cite{Bianzino2014} \\ 
\hline
\end{tabular}
\end{threeparttable}
\end{table}

\subsection{Non-functional properties}
\label{sec_nonfunctional}

In addition to enabling functionalities when using the edge computing paradigm the very organization of the edge architecture and realizing desirable properties requires some kind of resource management too. This additional work is not directly related to the service to obtain, i.e. it is a non-functional property (also referred to as extra-functional properties). Obviously, papers that are focusing on a functional property can also be interested in some non-functional property.

This subsection is related to the categories of objectives for resource management we have already discussed in section \ref{sec:Objective}.
Achieving the objectives in that section was evaluated using metrics that are often representative for measuring non-functional properties.

Examples of metrics and their related non-functional property that are encountered more often are:
\begin{itemize}
\item \textit{response time} as a measure of \textbf{timeliness}.
\item \textit{energy consumption} as a measure of \textbf{energy efficiency}.
\item \textit{admission ratio}, or its equivalent blocking probability, as a measure of \textbf{availability} of the edge service.
\item \textit{CPU/network utilization} as a measure of \textbf{computation/communication resource efficiency}.
\item \textit{Monetary cost} paid to an infrastructure owner as a measure of \textbf{cost efficiency}.
\end{itemize}

\begin{figure}[!t]
\centering
\includegraphics[width=3.5in]{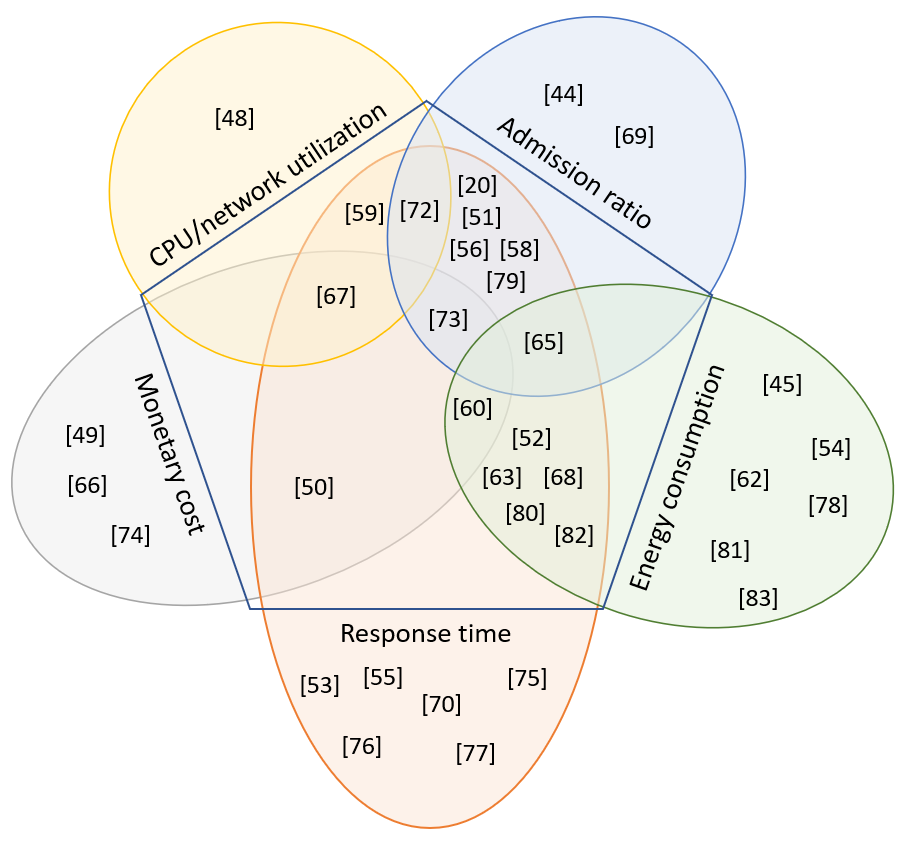}
\caption{Generic metrics related to a non-functional property used in the surveyed articles.}
\label{fig_NF}
\end{figure}

The list of metrics is not exhaustive, but we have focused on the more prevalent ones.
Figure \ref{fig_NF} shows how popular the above metrics are in the context of the works studied so far.

It is not surprising that those metrics that relate to timeliness or availability or resource efficiency are well-represented.

As we have noticed earlier, the same paper can deal with multiple resources, multiple objectives, and also clearly seen in this figure, multiple non-functional properties. This illustrates the complex trade-offs involved when dealing with resources in a multi-stakeholder distributed system.

\section{Research challenges}
\label{sec:Challenges}

In this section, we present the research challenges not substantially addressed which could be of interest for further research in the field.

From the previous sections, we noted that the architecture with three active and distinct levels (Edge server) is predominant. We also noted that the resource objectives allocation and optimization were well-studied. Moreover, computational and communicational resources are the most commonly addressed, typically being stationary and located within a single level. Therefore, research is less prevalent on data, storage, and energy as a resource, and less extensive towards the estimation, discovery and sharing objectives (especially the first two). Furthermore, new works should consider mobility and multi-level locality on the supply side.

Elaborating on mobility, the new phenomenon at the edge is that the supply side can also be mobile, and not only the demand side as it was the case in classic clouds. Indeed, edge systems will have to deal with a greater variety of mobility, with end devices that are often mobile (like vehicles) but can also be stationary (e.g. video-surveillance cameras), as well as mobile edge devices. It is however not obvious that the mobility patterns of all those devices will be similar, especially between end and edge devices. Considering the large variety of edge applications, their characteristics can potentially vary greatly. For example, an edge solution intended to serve networks of cars moving on a road network will probably be quite different from an edge solution intended to serve persons within a shopping mall. Hence, it is critical to have efficient, and thus tailored solutions. But should each application domain rediscover the wheel? Obviously, there is going to be generic wisdom that is transferable across the domains if adequate characterizations of resource requirement patterns are formulated. More work is needed on collecting mobility traces from the different edge applications to see if present patterns can in a generic way be used to create pertinent edge mobility models  at both levels of the architecture, the end and the edge level. These can then become a basis for repeatable evaluations of resource management strategies.

Another aspect which will be critical to solving is collaboration. There are new papers appearing where multiple operators at the edge level are modeled, and this introduces new challenges. At the end level, we have seen that different incentives can be provided to enable resource sharing \cite{DoubleSidedBidding,Bianzino2014}, and similarly at the edge level \cite{Chen_sociallyTrusted}. Such collaboration is especially good for managing workload churns and is interesting for infrastructure owners. The next challenge would be to do multi-level collaboration with a hierarchy of incentive schemes at different levels assuring that they do not cancel out each other's benefits. Moreover, finding more advanced incentive schemes  that take both resource efficiency and security into account is needed. Current solutions either choose to not collaborate for security- or privacy- sensitive tasks \cite{Singh_RTSANE} or rely on classic trust establishment \cite{Chen_sociallyTrusted} but this will not be enough for a wide collaboration at the edge.

Context adaptation is also one of the properties expected from edge computing and advocated as a good reason to choose this paradigm \cite{ContextSensitiveOffloading}. Providing tailored service depending on the user's physical location of course has to be taken care of at the application level. However, it also impacts resource management as those applications will require resources to provide those services, in particular considering data (about supply mobility and abundance) as a resource.

Security, and its subcomponents availability, confidentiality, and integrity, is a key point for edge computing, together with privacy with respect to sensitive end-user data. Although similar, security and privacy have distinct characteristics and should be addressed in depth and separately, which is not the case in the current surveyed works \cite{Singh_RTSANE,Chen_sociallyTrusted}. Regarding availability, most of the works considered focus on admission ratio but do not consider the fact that resources could disappear while executing due to mobility, misuses, or attacks. A notable exception is the work by Habak et al. \cite{Habak_WorkloadManagement} who propose and evaluate a task checkpointing mechanism which performs result replication to mitigate in case a device disappears. Focusing on availability, several works always consider that the cloud is available as a last fall-back for providing an edge service. If this is not the case (e.g. due to overload, attack, or natural disaster) the availability of the edge service will be impacted. More works in those directions and quantifying edge-specific availability metrics are required. Edge computing will most certainly be interesting for critical infrastructure because of its benefits and those require high standards on security. Research in this direction can be found for the mobile cloud paradigm, for example \cite{SecurityAwareResourceAllocation, ResourceAllocationSecurityMCC}, but they consider scenarios where the edge level is absent.

End-to-end timeliness requires quantification of latencies from an end device towards the cloud (or somewhere at the edge) all the way back to the same device (or to another device). This means traversing the edge networking services, including what we referred to as resource management services in this paper. Since estimation, discovery, sharing, allocation (including migration) are complex algorithms in such networks, these must also be evaluated in terms of their own resource footprint, and thereby their own impact on timeliness and QoS. In the surveyed articles, computing time of the solution is only evaluated by Gomes et al. \cite{EdgeCachingMobilityPrediction} and Skarlat et al. \cite{TowardsQoSAwareFogServicePlacement}. Since edge computing cannot become widely used without strong security and privacy properties, it is especially important to research on the resource overhead for providing those properties as well. Too high an overhead can signify a technology which is not feasible in practice.

As shown in Section \ref{sec:resourceType}, resources managed at the edge are most often a combination of different resource types. This implies that there will be some interrelations among resource utilization levels, which can create new challenges. Considering resource affinity as in the work by Yousaf et al. \cite{FinedGrainedResourceAware} may be a start but more research is needed to understand and address the complexity of such multi-resource problems in the edge context.

As mentioned in Section \ref{sec_architectureTaxonomy}, edge computing brings together diverse business sectors, with their existing techniques for solving relevant problems in those areas. Techniques previously applied in only one of those domains may be applicable to edge computing with the required adaptations. For example, performing resource migration requires efficient techniques for this purpose. Ma et al. \cite{Ma_ServiceHandoff} study container migration and found that the hand-off time decreased by 56\% to 80\% in comparison to state-of-the-art VM migration for the edge. Results like this should be exploited in the new edge era, and utilize technologies that may bring added benefit to edge computing.

Another enabler for resource-efficient edge computing is the development of tools for testing the new proposals in relevant conditions and setups. In the surveyed articles, the most common method used for validating a model or a proposed algorithm is to use an analytical tool (e.g. a solver and/or an optimization engine). Another common approach is to use a simulator, either a generic network simulator such as OMNeT++\footnote{\url{https://omnetpp.org/}}, or one designed for regular cloud environments such as CloudSim \cite{CloudSim}, most often with some custom extensions. There also exists a dedicated simulator designed for fog computing, called iFogSim \cite{iFogSimPublished}, which extends CloudSim, but this one currently has limitations, e.g. no mechanisms for offloading or communication between two nodes at the same level. A third way of evaluating in the surveyed works is the use of physical testbeds. Such evaluations provide invaluable insights into problems that are easy to oversee in simulation and investigate their impact. However, a big challenge for testbeds is to get them to scale, which is to some extent also a problem for simulations. Therefore, there is a need for creating open research testbeds and simulation tools so that configurable architectures and application/domain-specific edge computing methods can be efficiently compared. Coming back to a previous point, such tools should be able to handle mobility of end and edge devices and should obviously be scalable for evaluation of real-world scenarios.

\section{Conclusion}
\label{sec_conclusion}

The past decade has created tremendous expectations on IoT changing the landscape of data-driven services with benefits for multiple societal sectors. Many researchers have contributed to the development of technologies and addressed challenges that come with resource scarcity in the end devices. Other researchers with a background in cloud computing have looked at how to carry the data generated by the massive IoT deployments and how to efficiently use the cloud resources. The area of edge computing brings these two ends of the same service together in an emerging ecosystem and creates a means to discuss resource adequacy from an end-to-end perspective. In this paper we have tried to provide an overview, not from a cloud perspective, or an IoT device perspective, but with a focus on edge resource management.

\section*{Acknowledgments}
This work was supported by the Swedish national graduate
school in computer science (CUGS).

\subsection*{Conflict of interest}
The authors declare that there is no conflict of interest regarding the publication of this paper.

\bibliographystyle{IEEEtran}
\bibliography{refs}

\begin{thebibliography}{10}
\providecommand{\url}[1]{#1}
\csname url@samestyle\endcsname
\providecommand{\newblock}{\relax}
\providecommand{\bibinfo}[2]{#2}
\providecommand{\BIBentrySTDinterwordspacing}{\spaceskip=0pt\relax}
\providecommand{\BIBentryALTinterwordstretchfactor}{4}
\providecommand{\BIBentryALTinterwordspacing}{\spaceskip=\fontdimen2\font plus
\BIBentryALTinterwordstretchfactor\fontdimen3\font minus
  \fontdimen4\font\relax}
\providecommand{\BIBforeignlanguage}[2]{{%
\expandafter\ifx\csname l@#1\endcsname\relax
\typeout{** WARNING: IEEEtran.bst: No hyphenation pattern has been}%
\typeout{** loaded for the language `#1'. Using the pattern for}%
\typeout{** the default language instead.}%
\else
\language=\csname l@#1\endcsname
\fi
#2}}
\providecommand{\BIBdecl}{\relax}
\BIBdecl

\bibitem{EricssonMobilityReport2017}
\BIBentryALTinterwordspacing
Ericsson. (2017, June) Ericsson mobility report. [Online]. Available:
  \url{https://www.ericsson.com/assets/local/mobility-report/documents/2017/ericsson-mobility-report-june-2017.pdf}
\BIBentrySTDinterwordspacing

\bibitem{EdgeAnalyticsIoT}
M.~Satyanarayanan, P.~Simoens, Y.~Xiao, P.~Pillai, Z.~Chen, K.~Ha, W.~Hu, and
  B.~Amos, ``Edge analytics in the internet of things,'' \emph{IEEE Pervasive
  Computing}, vol.~14, no.~2, pp. 24--31, Apr 2015.

\bibitem{EmergenceEdge}
M.~Satyanarayanan, ``The emergence of edge computing,'' \emph{Computer},
  vol.~50, no.~1, pp. 30--39, Jan 2017.

\bibitem{HowBeneficialIntermediateDC}
A.~Mehta, W.~Tärneberg, C.~Klein, J.~Tordsson, M.~Kihl, and E.~Elmroth, ``How
  beneficial are intermediate layer data centers in mobile edge networks?'' in
  \emph{2016 IEEE 1st International Workshops on Foundations and Applications
  of Self* Systems (FAS*W)}, Sept 2016, pp. 222--229.

\bibitem{DEVS}
M.~Etemad, M.~Aazam, and M.~St-Hilaire, ``Using {DEVS} for modeling and
  simulating a fog computing environment,'' in \emph{2017 International
  Conference on Computing, Networking and Communications (ICNC)}, Jan 2017, pp.
  849--854.

\bibitem{FernandoSurvey}
N.~Fernando, S.~W. Loke, and W.~Rahayu, ``Mobile cloud computing: A survey,''
  \emph{Future Generation Computer Systems}, vol.~29, no.~1, pp. 84 -- 106,
  2013.

\bibitem{FogReferencePaper}
F.~Bonomi, R.~Milito, J.~Zhu, and S.~Addepalli, ``Fog computing and its role in
  the internet of things,'' in \emph{Proceedings of the First Edition of the
  MCC Workshop on Mobile Cloud Computing (MCC '12)}.\hskip 1em plus 0.5em minus
  0.4em\relax New York, NY, USA: ACM, 2012, pp. 13--16.

\bibitem{RomanSurvey}
R.~Roman, J.~Lopez, and M.~Mambo, ``Mobile edge computing, fog et al.: A survey
  and analysis of security threats and challenges,'' \emph{Future Generation
  Computer Systems}, vol.~78, no. Part 2, pp. 680 -- 698, 2018.

\bibitem{Mortazavi_CloudPath}
S.~H. Mortazavi, M.~Salehe, C.~S. Gomes, C.~Phillips, and E.~de~Lara,
  ``Cloudpath: A multi-tier cloud computing framework,'' in \emph{Proceedings
  of the Second ACM/IEEE Symposium on Edge Computing}, ser. SEC '17.\hskip 1em
  plus 0.5em minus 0.4em\relax New York, NY, USA: ACM, 2017, pp. 20:1--20:13.

\bibitem{LiuSurvey}
H.~Liu, F.~Eldarrat, H.~Alqahtani, A.~Reznik, X.~de~Foy, and Y.~Zhang, ``Mobile
  edge cloud system: Architectures, challenges, and approaches,'' \emph{IEEE
  Systems Journal}, 2017, {DOI}: 10.1109/JSYST.2017.2654119.

\bibitem{TarnebergResourceManagementChallenges}
W.~Tärneberg, A.~Mehta, J.~Tordsson, M.~Kihl, and E.~Elmroth, ``Resource
  management challenges for the infinite cloud,'' in \emph{Paper presented at
  10th International Workshop on Feedback Computing at CPSWeek 2017}, Seattle,
  United States, 2015.

\bibitem{Taleb_FollowMeCloud}
T.~Taleb and A.~Ksentini, ``Follow me cloud: interworking federated clouds and
  distributed mobile networks,'' \emph{IEEE Network}, vol.~27, no.~5, pp.
  12--19, September 2013.

\bibitem{EdgeCachingMobilityPrediction}
A.~S. Gomes, B.~Sousa, D.~Palma, V.~Fonseca, Z.~Zhao, E.~Monteiro, T.~Braun,
  P.~Simoes, and L.~Cordeiro, ``Edge caching with mobility prediction in
  virtualized {LTE} mobile networks,'' \emph{Future Generation Computer
  Systems}, vol.~70, pp. 148 -- 162, 2017.

\bibitem{MultiCloudThesisVamisPaperChapter6}
V.~Xhagjika, L.~Navarro, and V.~Vlassov, ``Enhancing real-time applications by
  means of multi-tier cloud federations,'' in \emph{2015 IEEE 7th International
  Conference on Cloud Computing Technology and Science (CloudCom)}, Nov 2015,
  pp. 397--404.

\bibitem{Lobillo_SmallCellCloud}
F.~Lobillo, Z.~Becvar, M.~A. Puente, P.~Mach, F.~L. Presti, F.~Gambetti,
  M.~Goldhamer, J.~Vidal, A.~K. Widiawan, and E.~Calvanesse, ``An architecture
  for mobile computation offloading on cloud-enabled {LTE} small cells,'' in
  \emph{2014 IEEE Wireless Communications and Networking Conference Workshops
  (WCNCW)}, April 2014, pp. 1--6.

\bibitem{Wang_FastMovingPersonalCloud}
K.~Wang, M.~Shen, J.~Cho, A.~Banerjee, J.~Van~der Merwe, and K.~Webb,
  ``Mobiscud: A fast moving personal cloud in the mobile network,'' in
  \emph{Proceedings of the 5th Workshop on All Things Cellular: Operations,
  Applications and Challenges}, ser. AllThingsCellular '15.\hskip 1em plus
  0.5em minus 0.4em\relax New York, NY, USA: ACM, 2015, pp. 19--24.

\bibitem{Liu_CONCERT}
J.~Liu, T.~Zhao, S.~Zhou, Y.~Cheng, and Z.~Niu, ``Concert: a cloud-based
  architecture for next-generation cellular systems,'' \emph{IEEE Wireless
  Communications}, vol.~21, no.~6, pp. 14--22, December 2014.

\bibitem{DistributedCloudSurvey}
P.~T. Endo, A.~V. de~Almeida~Palhares, N.~N. Pereira, G.~E. Goncalves,
  D.~Sadok, J.~Kelner, B.~Melander, and J.~E. Mangs, ``Resource allocation for
  distributed cloud: concepts and research challenges,'' \emph{IEEE Network},
  vol.~25, no.~4, pp. 42--46, July 2011.

\bibitem{Habak_FemtoCloud}
K.~Habak, M.~Ammar, K.~A. Harras, and E.~Zegura, ``Femto clouds: Leveraging
  mobile devices to provide cloud service at the edge,'' in \emph{2015 IEEE 8th
  International Conference on Cloud Computing}, June 2015, pp. 9--16.

\bibitem{Habak_WorkloadManagement}
K.~Habak, E.~W. Zegura, M.~Ammar, and K.~A. Harras, ``Workload management for
  dynamic mobile device clusters in edge femtoclouds,'' in \emph{Proceedings of
  the Second ACM/IEEE Symposium on Edge Computing}, ser. SEC '17.\hskip 1em
  plus 0.5em minus 0.4em\relax New York, NY, USA: ACM, 2017, pp. 6:1--6:14.

\bibitem{AhmedSurvey}
A.~Ahmed and E.~Ahmed, ``A survey on mobile edge computing,'' in \emph{2016
  10th International Conference on Intelligent Systems and Control (ISCO)}, Jan
  2016, pp. 1--8.

\bibitem{StojmenovicSurvey}
I.~Stojmenovic, ``Fog computing: A cloud to the ground support for smart things
  and machine-to-machine networks,'' in \emph{2014 Australasian
  Telecommunication Networks and Applications Conference (ATNAC)}, Nov 2014,
  pp. 117--122.

\bibitem{ChiangSurvey}
M.~Chiang and T.~Zhang, ``Fog and {IoT}: An overview of research
  opportunities,'' \emph{IEEE Internet of Things Journal}, vol.~3, no.~6, pp.
  854--864, Dec 2016.

\bibitem{WeisongEdgeVisionChallenges}
W.~Shi, J.~Cao, Q.~Zhang, Y.~Li, and L.~Xu, ``Edge computing: Vision and
  challenges,'' \emph{IEEE Internet of Things Journal}, vol.~3, no.~5, pp.
  637--646, Oct 2016.

\bibitem{DataAnalyticsNetworkEdge}
N.~M. Gonzalez, W.~A. Goya, R.~de~Fatima~Pereira, K.~Langona, E.~A. Silva,
  T.~C.~M. de~Brito~Carvalho, C.~C. Miers, J.~E. Mångs, and A.~Sefidcon, ``Fog
  computing: Data analytics and cloud distributed processing on the network
  edges,'' in \emph{2016 35th International Conference of the Chilean Computer
  Science Society (SCCC)}, Oct 2016, pp. 1--9.

\bibitem{QunSecurityPrivacyFogSurvey}
S.~Yi, Z.~Qin, and Q.~Li, ``Security and privacy issues of fog computing: A
  survey,'' in \emph{Wireless Algorithms, Systems, and Applications: 10th
  International Conference}, August 2015, pp. 685--695.

\bibitem{Baktir_Survey}
A.~C. Baktir, A.~Ozgovde, and C.~Ersoy, ``How can edge computing benefit from
  software-defined networking: A survey, use cases, and future directions,''
  \emph{IEEE Communications Surveys Tutorials}, vol.~19, no.~4, pp. 2359--2391,
  Fourthquarter 2017.

\bibitem{YiSurvey}
S.~Yi, C.~Li, and Q.~Li, ``A survey of fog computing: Concepts, applications
  and issues,'' in \emph{Proceedings of the 2015 Workshop on Mobile Big Data
  (Mobidata '15)}.\hskip 1em plus 0.5em minus 0.4em\relax New York, NY, USA:
  ACM, 2015, pp. 37--42.

\bibitem{MahmudSurvey}
R.~Mahmud, K.~Ramamohanarao, and R.~Buyya, ``Fog computing: {A} taxonomy,
  survey and future directions,'' \emph{Internet of Everything. Internet of
  Things (Technology, Communications and Computing)}, 2018.

\bibitem{OffloadingSurvey}
A.~Bhattacharya and P.~De, ``A survey of adaptation techniques in computation
  offloading,'' \emph{Journal of Network and Computer Applications}, vol.~78,
  pp. 97 -- 115, 2017.

\bibitem{OffloadingSurvey2}
F.~Rebecchi, M.~D. de~Amorim, V.~Conan, A.~Passarella, R.~Bruno, and M.~Conti,
  ``Data offloading techniques in cellular networks: A survey,'' \emph{IEEE
  Communications Surveys Tutorials}, vol.~17, no.~2, pp. 580--603, 2015.

\bibitem{MaoSurveyUnpublished}
Y.~{Mao}, C.~{You}, J.~{Zhang}, K.~{Huang}, and K.~B. {Letaief}, ``{Mobile Edge
  Computing: Survey and Research Outlook},'' Jan. 2017, preprint,
  arXiv:1701.01090.

\bibitem{MachSurvey}
P.~Mach and Z.~Becvar, ``Mobile edge computing: A survey on architecture and
  computation offloading,'' \emph{IEEE Communications Surveys Tutorials},
  vol.~19, no.~3, pp. 1628--1656, thirdquarter 2017.

\bibitem{MaoSurveyPublished}
Y.~Mao, C.~You, J.~Zhang, K.~Huang, and K.~B. Letaief, ``A survey on mobile
  edge computing: The communication perspective,'' \emph{IEEE Communications
  Surveys Tutorials}, vol.~19, no.~4, pp. 2322--2358, Fourthquarter 2017.

\bibitem{ChunlinCostEnergyAware}
L.~Chunlin and L.~LaYuan, ``Cost and energy aware service provisioning for
  mobile client in cloud computing environment,'' \emph{The Journal of
  Supercomputing}, vol.~71, no.~4, pp. 1196--1223, 2015.

\bibitem{Indices}
S.~Shekhar, A.~D. Chhokra, A.~Bhattacharjee, G.~Aupy, and A.~Gokhale,
  ``{INDICES: E}xploiting edge resources for performance-aware cloud-hosted
  services,'' in \emph{2017 IEEE 1st International Conference on Fog and Edge
  Computing (ICFEC)}, May 2017, pp. 75--80.

\bibitem{SecureEdgePaper}
K.~Toczé and S.~Nadjm-Tehrani, ``Where resources meet at the edge,'' in
  \emph{2017 IEEE International Conference on Computer and Information
  Technology (CIT)}, Aug 2017, pp. 302--307.

\bibitem{RichardContainer}
R.~Cziva and D.~P. Pezaros, ``Container network functions: Bringing {NFV} to
  the network edge,'' \emph{IEEE Communications Magazine}, vol.~55, no.~6, pp.
  24--31, 2017.

\bibitem{MouradianSurvey}
C.~Mouradian, D.~Naboulsi, S.~Yangui, R.~H. Glitho, M.~J. Morrow, and P.~A.
  Polakos, ``A comprehensive survey on fog computing: State-of-the-art and
  research challenges,'' \emph{IEEE Communications Surveys Tutorials}, 2017,
  {DOI}: 10.1109/COMST.2017.2771153.

\bibitem{TowardsMobileOpportunisticComputing}
A.~Mtibaa, K.~A. Harras, K.~Habak, M.~Ammar, and E.~W. Zegura, ``Towards mobile
  opportunistic computing,'' in \emph{2015 IEEE 8th International Conference on
  Cloud Computing}, June 2015, pp. 1111--1114.

\bibitem{CloudletReferencePaper}
M.~Satyanarayanan, P.~Bahl, R.~Caceres, and N.~Davies, ``The case for
  {VM}-based cloudlets in mobile computing,'' \emph{IEEE Pervasive Computing},
  vol.~8, no.~4, pp. 14--23, Oct 2009.

\bibitem{CSCAN}
H.~Frank, W.~F. Fuhrmann, and B.~V. Ghita, ``Mobile edge computing:
  Requirements for powerful mobile near real-time applications,'' in
  \emph{Proceedings of the Eleventh International Network Conference (INC
  2016)}.\hskip 1em plus 0.5em minus 0.4em\relax Plymouth University, 2016, pp.
  63--66.

\bibitem{PREFog}
M.~Aazam, M.~St-Hilaire, C.~H. Lung, and I.~Lambadaris, ``{PRE-Fog: IoT} trace
  based probabilistic resource estimation at fog,'' in \emph{2016 13th IEEE
  Annual Consumer Communications Networking Conference (CCNC)}, Jan 2016, pp.
  12--17.

\bibitem{Singh_RTSANE}
A.~Singh, N.~Auluck, O.~Rana, A.~Jones, and S.~Nepal, ``{RT-SANE}: Real time
  security aware scheduling on the network edge,'' in \emph{Proceedings of
  the10th International Conference on Utility and Cloud Computing}, ser. UCC
  '17.\hskip 1em plus 0.5em minus 0.4em\relax New York, NY, USA: ACM, 2017, pp.
  131--140.

\bibitem{GreeningNanoDataCenters}
V.~Valancius, N.~Laoutaris, L.~Massouli{\'e}, C.~Diot, and P.~Rodriguez,
  ``Greening the internet with nano data centers,'' in \emph{Proceedings of the
  5th International Conference on Emerging Networking Experiments and
  Technologies}, ser. CoNEXT '09.\hskip 1em plus 0.5em minus 0.4em\relax New
  York, NY, USA: ACM, 2009, pp. 37--48.

\bibitem{ResourecAllocSchemeVehicular}
H.~Meng, K.~Zheng, P.~Chatzimisios, H.~Zhao, and L.~Ma, ``A utility-based
  resource allocation scheme in cloud-assisted vehicular network
  architecture,'' in \emph{2015 IEEE International Conference on Communication
  Workshop (ICCW)}, June 2015, pp. 1833--1838.

\bibitem{5GSurvey}
M.~Agiwal, A.~Roy, and N.~Saxena, ``Next generation {5G} wireless networks: A
  comprehensive survey,'' \emph{IEEE Communications Surveys Tutorials},
  vol.~18, no.~3, pp. 1617--1655, 2016.

\bibitem{FinedGrainedResourceAware}
F.~Z. Yousaf and T.~Taleb, ``Fine-grained resource-aware virtual network
  function management for {5G} carrier cloud,'' \emph{IEEE Network}, vol.~30,
  no.~2, pp. 110--115, March 2016.

\bibitem{CostEffectiveResourceAllocaion}
K.~Wang, K.~Yang, X.~Wang, and C.~S. Magurawalage, ``Cost-effective resource
  allocation in {C-RAN} with mobile cloud,'' in \emph{2016 IEEE International
  Conference on Communications (ICC)}, May 2016, pp. 1--6.

\bibitem{TowardsQoSAwareFogServicePlacement}
O.~Skarlat, M.~Nardelli, S.~Schulte, and S.~Dustdar, ``Towards {QoS}-aware fog
  service placement,'' in \emph{2017 IEEE 1st International Conference on Fog
  and Edge Computing (ICFEC)}, May 2017, pp. 89--96.

\bibitem{ClusterBasedVehicularClouds}
H.~R. Arkian, R.~E. Atani, and A.~Pourkhalili, ``A cluster-based vehicular
  cloud architecture with learning-based resource management,'' in \emph{2014
  IEEE 6th International Conference on Cloud Computing Technology and Science},
  Dec 2014, pp. 162--167.

\bibitem{ServiceOrientedHeterogeneousResourceSharing}
T.~Nishio, R.~Shinkuma, T.~Takahashi, and N.~B. Mandayam, ``Service-oriented
  heterogeneous resource sharing for optimizing service latency in mobile
  cloud,'' in \emph{Proceedings of the First International Workshop on Mobile
  Cloud Computing \&\#38; Networking}, ser. MobileCloud '13.\hskip 1em plus
  0.5em minus 0.4em\relax New York, NY, USA: ACM, 2013, pp. 19--26.

\bibitem{ServiceProvisioningOpportunistic}
D.~Mascitti, M.~Conti, A.~Passarella, and L.~Ricci, ``Service provisioning
  through opportunistic computing in mobile clouds,'' vol.~40, no. Supplement
  C, pp. 143 -- 150, 2014, fourth International Conference on Selected Topics
  in Mobile and Wireless Networking (MoWNet’2014).

\bibitem{AdaptiveResourceDiscovery}
W.~Liu, T.~Nishio, R.~Shinkuma, and T.~Takahashi, ``Adaptive resource discovery
  in mobile cloud computing,'' \emph{Comput. Commun.}, vol.~50, pp. 119--129,
  Sep. 2014.

\bibitem{OpportunisticResourceSharingMCC}
W.~Liu, R.~Shinkuma, and T.~Takahashi, ``Opportunistic resource sharing in
  mobile cloud computing: The single-copy case,'' in \emph{The 16th
  Asia-Pacific Network Operations and Management Symposium}, Sept 2014, pp.
  1--6.

\bibitem{TransientCloud}
T.~Penner, A.~Johnson, B.~V. Slyke, M.~Guirguis, and Q.~Gu, ``Transient clouds:
  Assignment and collaborative execution of tasks on mobile devices,'' in
  \emph{2014 IEEE Global Communications Conference}, Dec 2014, pp. 2801--2806.

\bibitem{WMFOG}
Z.~Hao, E.~Novak, S.~Yi, and Q.~Li, ``Challenges and software architecture for
  fog computing,'' \emph{IEEE Internet Computing}, vol.~21, no.~2, pp. 44--53,
  Mar 2017.

\bibitem{AdaptiveMultiResourceAllocationCloudlet}
Y.~Liu, M.~J. Lee, and Y.~Zheng, ``Adaptive multi-resource allocation for
  cloudlet-based mobile cloud computing system,'' \emph{IEEE Transactions on
  Mobile Computing}, vol.~15, no.~10, pp. 2398--2410, Oct 2016.

\bibitem{ObjectStore}
B.~Confais, A.~Lebre, and B.~Parrein, ``An object store service for a fog/edge
  computing infrastructure based on {IPFS} and a scale-out {NAS},'' in
  \emph{2017 IEEE 1st International Conference on Fog and Edge Computing
  (ICFEC)}, May 2017, pp. 41--50.

\bibitem{MIST}
H.~R. Arkian, A.~Diyanat, and A.~Pourkhalili, ``{MIST}: Fog-based data
  analytics scheme with cost-efficient resource provisioning for {IoT}
  crowdsensing applications,'' \emph{Journal of Network and Computer
  Applications}, vol.~82, pp. 152 -- 165, 2017.

\bibitem{FogDynamicResourceEstimation}
M.~Aazam and E.~N. Huh, ``Fog computing micro datacenter based dynamic resource
  estimation and pricing model for {IoT},'' in \emph{2015 IEEE 29th
  International Conference on Advanced Information Networking and
  Applications}, March 2015, pp. 687--694.

\bibitem{EnergyDrivenAvatar}
Q.~Fan, N.~Ansari, and X.~Sun, ``Energy driven avatar migration in green
  cloudlet networks,'' \emph{IEEE Communications Letters}, vol.~21, no.~7, pp.
  1601--1604, July 2017.

\bibitem{OueisThesis}
J.~Oueis, ``Joint communication and computation resources allocation for
  cloud-empowered future wireless networks,'' Ph.D. dissertation, Université
  Grenoble Alpes, 2016.

\bibitem{DoubleSidedBidding}
L.~Tang, S.~He, and Q.~Li, ``Double-sided bidding mechanism for resource
  sharing in mobile cloud,'' \emph{IEEE Transactions on Vehicular Technology},
  vol.~66, no.~2, pp. 1798--1809, Feb 2017.

\bibitem{EnergyAwareFogAndCloud}
P.~Borylo, A.~Lason, J.~Rzasa, A.~Szymanski, and A.~Jajszczyk, ``Energy-aware
  fog and cloud interplay supported by wide area software defined networking,''
  in \emph{2016 IEEE International Conference on Communications (ICC)}, May
  2016, pp. 1--7.

\bibitem{CostEfficient}
L.~Gu, D.~Zeng, S.~Guo, A.~Barnawi, and Y.~Xiang, ``Cost efficient resource
  management in fog computing supported medical cyber-physical system,''
  \emph{IEEE Transactions on Emerging Topics in Computing}, vol.~5, no.~1, pp.
  108--119, Jan 2017.

\bibitem{DynamicApplicationPlacement}
W.~Tärneberg, A.~Mehta, E.~Wadbro, J.~Tordsson, J.~Eker, M.~Kihl, and
  E.~Elmroth, ``Dynamic application placement in the mobile cloud network,''
  \emph{Future Generation Computer Systems}, vol.~70, pp. 163 -- 177, 2017.

\bibitem{DynamicResourceAlllocation}
J.~Plachy, Z.~Becvar, and E.~C. Strinati, ``Dynamic resource allocation
  exploiting mobility prediction in mobile edge computing,'' in \emph{2016 IEEE
  27th Annual International Symposium on Personal, Indoor, and Mobile Radio
  Communications (PIMRC)}, Sept 2016, pp. 1--6.

\bibitem{OffloadingSchemeDCFog}
C.~Fricker, F.~Guillemin, P.~Robert, and G.~Thompson, ``Analysis of an
  offloading scheme for data centers in the framework of fog computing,''
  \emph{ACM Trans. Model. Perform. Eval. Comput. Syst.}, vol.~1, no.~4, pp.
  16:1--16:18, Sep. 2016.

\bibitem{HybridMethodServiceDelay}
T.~G. Rodrigues, K.~Suto, H.~Nishiyama, and N.~Kato, ``Hybrid method for
  minimizing service delay in edge cloud computing through vm migration and
  transmission power control,'' \emph{IEEE Transactions on Computers}, vol.~66,
  no.~5, pp. 810--819, May 2017.

\bibitem{WeisongFirework}
Q.~Zhang, X.~Zhang, Q.~Zhang, W.~Shi, and H.~Zhong, ``Firework: Big data
  sharing and processing in collaborative edge environment,'' in \emph{2016
  Fourth IEEE Workshop on Hot Topics in Web Systems and Technologies (HotWeb)},
  Oct 2016, pp. 20--25.

\bibitem{MobilityAwareApplicationScheduling}
L.~F. Bittencourt, J.~Diaz-Montes, R.~Buyya, O.~F. Rana, and M.~Parashar,
  ``Mobility-aware application scheduling in fog computing,'' \emph{IEEE Cloud
  Computing}, vol.~4, no.~2, pp. 26--35, March 2017.

\bibitem{DeadlineConstrainedVideoAnalysis}
A.~R. Zamani, M.~Zou, J.~Diaz-Montes, I.~Petri, O.~Rana, A.~Anjum, and
  M.~Parashar, ``Deadline constrained video analysis via in-transit
  computational environments,'' \emph{IEEE Transactions on Services Computing},
  2017, {DOI}: 10.1109/TSC.2017.2653116.

\bibitem{Chen_sociallyTrusted}
L.~Chen and J.~Xu, ``Socially trusted collaborative edge computing in ultra
  dense networks,'' in \emph{Proceedings of the Second ACM/IEEE Symposium on
  Edge Computing}, ser. SEC '17.\hskip 1em plus 0.5em minus 0.4em\relax New
  York, NY, USA: ACM, 2017, pp. 9:1--9:11.

\bibitem{Wang_ElasticUrbanVideosur}
J.~Wang, J.~Pan, and F.~Esposito, ``Elastic urban video surveillance system
  using edge computing,'' in \emph{Proceedings of the Workshop on Smart
  Internet of Things}, ser. SmartIoT '17.\hskip 1em plus 0.5em minus
  0.4em\relax New York, NY, USA: ACM, 2017, pp. 7:1--7:6.

\bibitem{Yi_LAVEA}
S.~Yi, Z.~Hao, Q.~Zhang, Q.~Zhang, W.~Shi, and Q.~Li, ``Lavea: Latency-aware
  video analytics on edge computing platform,'' in \emph{Proceedings of the
  Second ACM/IEEE Symposium on Edge Computing}, ser. SEC '17.\hskip 1em plus
  0.5em minus 0.4em\relax New York, NY, USA: ACM, 2017, pp. 15:1--15:13.

\bibitem{Wang_DynamicServicePlacementJournal}
S.~Wang, R.~Urgaonkar, T.~He, K.~Chan, M.~Zafer, and K.~K. Leung, ``Dynamic
  service placement for mobile micro-clouds with predicted future costs,''
  \emph{IEEE Transactions on Parallel and Distributed Systems}, vol.~28, no.~4,
  pp. 1002--1016, April 2017.

\bibitem{Sardellitti_JointOptimization}
S.~Sardellitti, G.~Scutari, and S.~Barbarossa, ``Joint optimization of radio
  and computational resources for multicell mobile-edge computing,'' \emph{IEEE
  Transactions on Signal and Information Processing over Networks}, vol.~1,
  no.~2, pp. 89--103, June 2015.

\bibitem{MobileEdgeCloudsForInformationCentric}
E.~Borgia, R.~Bruno, M.~Conti, D.~Mascitti, and A.~Passarella, ``Mobile edge
  clouds for information-centric {IoT} services,'' in \emph{2016 IEEE Symposium
  on Computers and Communication (ISCC)}, June 2016, pp. 422--428.

\bibitem{ResourceDiscoveryInMCC}
P.~Athwani and D.~P. Vidyarthi, ``Resource discovery in mobile cloud computing:
  A clustering based approach,'' in \emph{2015 IEEE UP Section Conference on
  Electrical Computer and Electronics (UPCON)}, Dec 2015, pp. 1--6.

\bibitem{Bianzino2014}
A.~P. Bianzino, J.-L. Rougier, C.~Chaudet, and D.~Rossi, ``The green-game:
  Accounting for device criticality in resource consolidation for backbone {IP}
  networks,'' \emph{Strategic Behavior and the Environment}, vol.~4, no.~2, pp.
  131--153, 2014.

\bibitem{DynamicResourceOrchestration}
Q.~Qi, J.~Liao, J.~Wang, Q.~Li, and Y.~Cao, ``Dynamic resource orchestration
  for multi-task application in heterogeneous mobile cloud computing,'' in
  \emph{2016 IEEE Conference on Computer Communications Workshops (INFOCOM
  WKSHPS)}, April 2016, pp. 221--226.

\bibitem{TowardsResourceSharing}
A.~Mtibaa, A.~Fahim, K.~A. Harras, and M.~H. Ammar, ``Towards resource sharing
  in mobile device clouds: Power balancing across mobile devices,'' in
  \emph{Proceedings of the Second ACM SIGCOMM Workshop on Mobile Cloud
  Computing}, ser. MCC '13.\hskip 1em plus 0.5em minus 0.4em\relax New York,
  NY, USA: ACM, 2013, pp. 51--56.

\bibitem{TOMPECS16}
E.~Vergara, S.~Nadjm-Tehrani, and M.~Asplund, ``Fairness and incentive
  considerations in energy apportionment policies,'' \emph{ACM Transactions on
  Modelling and Performance Evaluation of Computer Systems (TOMPECS)}, 2016.

\bibitem{Xiao2013}
Z.~Xiao, W.~Song, and Q.~Chen, ``Dynamic resource allocation using virtual
  machines for cloud computing environment,'' \emph{IEEE Transactions on
  Parallel and Distributed Systems}, vol.~24, no.~6, pp. 1107 -- 1117, 2013.

\bibitem{Dong_EnergyEfficient}
Z.~Dong, Y.~Liu, H.~Zhou, X.~Xiao, Y.~Gu, L.~Zhang, and C.~Liu, ``An
  energy-efficient offloading framework with predictable temporal
  correctness,'' in \emph{Proceedings of the Second ACM/IEEE Symposium on Edge
  Computing}, ser. SEC '17.\hskip 1em plus 0.5em minus 0.4em\relax New York,
  NY, USA: ACM, 2017, pp. 19:1--19:12.

\bibitem{CooperativeResourceManagementVehicular}
R.~Yu, X.~Huang, J.~Kang, J.~Ding, S.~Maharjan, S.~Gjessing, and Y.~Zhang,
  ``Cooperative resource management in cloud-enabled vehicular networks,''
  \emph{IEEE Transactions on Industrial Electronics}, vol.~62, no.~12, pp.
  7938--7951, Dec 2015.

\bibitem{ENORM}
N.~Wang, B.~Varghese, M.~Matthaiou, and D.~S. Nikolopoulos, ``{ENORM}: A
  framework for edge node resource management,'' \emph{IEEE Transactions on
  Services Computing}, 2017, {DOI}: 10.1109/TSC.2017.2753775.

\bibitem{Satya2013}
M.~Satyanarayanan, G.~Lewis, E.~Morris, S.~Simanta, J.~Boleng, and K.~Ha, ``The
  role of cloudlets in hostile environments,'' \emph{IEEE Pervasive Computing},
  vol.~12, no.~4, pp. 40--49, Oct 2013.

\bibitem{ContextSensitiveOffloading}
B.~Zhou, A.~V. Dastjerdi, R.~N. Calheiros, S.~N. Srirama, and R.~Buyya, ``A
  context sensitive offloading scheme for mobile cloud computing service,'' in
  \emph{2015 IEEE 8th International Conference on Cloud Computing}, June 2015,
  pp. 869--876.

\bibitem{SecurityAwareResourceAllocation}
Y.~Liu and M.~J. Lee, ``Security-aware resource allocation for mobile cloud
  computing systems,'' in \emph{2015 24th International Conference on Computer
  Communication and Networks (ICCCN)}, Aug 2015, pp. 1--8.

\bibitem{ResourceAllocationSecurityMCC}
H.~Liang, D.~Huang, L.~X. Cai, X.~Shen, and D.~Peng, ``Resource allocation for
  security services in mobile cloud computing,'' in \emph{2011 IEEE Conference
  on Computer Communications Workshops (INFOCOM WKSHPS)}, April 2011, pp.
  191--195.

\bibitem{Ma_ServiceHandoff}
L.~Ma, S.~Yi, and Q.~Li, ``Efficient service handoff across edge servers via
  docker container migration,'' in \emph{Proceedings of the Second ACM/IEEE
  Symposium on Edge Computing}, ser. SEC '17.\hskip 1em plus 0.5em minus
  0.4em\relax New York, NY, USA: ACM, 2017, pp. 11:1--11:13.

\bibitem{CloudSim}
R.~N. Calheiros, R.~Ranjan, A.~Beloglazov, C.~A.~F. De~Rose, and R.~Buyya,
  ``{CloudSim}: A toolkit for modeling and simulation of cloud computing
  environments and evaluation of resource provisioning algorithms,''
  \emph{Softw. Pract. Exper.}, vol.~41, no.~1, pp. 23--50, Jan. 2011.

\bibitem{iFogSimPublished}
\BIBentryALTinterwordspacing
H.~Gupta, A.~Vahid~Dastjerdi, S.~K. Ghosh, and R.~Buyya, ``ifogsim: A toolkit
  for modeling and simulation of resource management techniques in the internet
  of things, edge and fog computing environments,'' \emph{Software: Practice
  and Experience}, vol.~47, no.~9, pp. 1275--1296, 2017, spe.2509. [Online].
  Available: \url{http://dx.doi.org/10.1002/spe.2509}
\BIBentrySTDinterwordspacing

\end{thebibliography}

\begin{IEEEbiography}[{\includegraphics[width=1in,height=1.25in,clip,keepaspectratio]{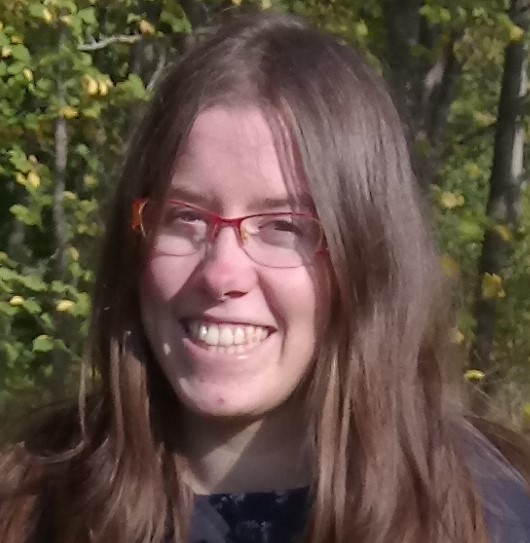}}]{Klervie Toczé}
Klervie Toczé received her French engineering degree from the University of Technology, Compiègne, France in 2014, and her M. Sc. degree in Computer Science from Linköping University, Sweden, in 2015. She is now pursuing Ph.D. studies at the Department of Computer and Information Science at Linköping University within the real-time Systems Laboratory. Her research interests are about resource management within edge/fog computing. 
\end{IEEEbiography}

\begin{IEEEbiography}[{\includegraphics[width=1in,height=1.25in,clip,keepaspectratio]{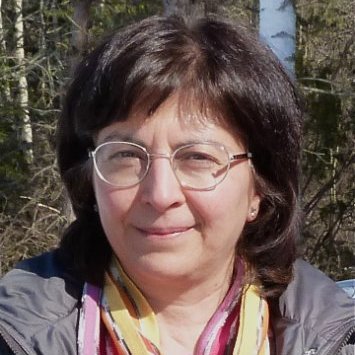}}]{Simin Nadjm-Tehrani}
Simin Nadjm-Tehrani received her B.Sc.
degree from Manchester University, UK, and
did her postgraduate studies leading to a Ph.D.
in Computer Science at Linköping University,
Sweden, in 1994. During 2006–2008 she was
a full professor at University of Luxembourg,
and is currently a Professor in Dependable
Distributed Systems at Department of Computer and Information Science, Linköping
University, where she has led the Real-time
Systems Laboratory since 2000. Her research
interests relate to networks and systems with
dependability requirements and resource constraints. 
\end{IEEEbiography}

\end{document}